\documentclass{article}
\usepackage{amsfonts}
\usepackage{amsmath}
\usepackage{amssymb}
\usepackage{graphicx}
\begin{document}

\title{Astrophysical black holes may radiate \\
but they do not evaporate}
\author{George F R Ellis \\
\\
\emph{Mathematics Department and ACGC, University of Cape Town};\\
\emph{Trinity College and DAMTP, Cambridge University.}
}
\maketitle

\begin{abstract}
\emph{This paper argues that the effect of Hawking radiation on an
astrophysical black hole situated in a realistic cosmological context is not total evaporation of the black hole; rather there will always be a remnant mass. The key point is that the locus of
emission of Hawking radiation is not the globally defined event horizon. Rather the emission domain lies just
outside a timelike Marginal Outer Trapped Surface that is locally defined. The emission domain is mainly located inside the event horizon. A spacelike singularity forms behind the event horizon, and most of the Hawking radiation ends up at this singularity rather than at infinity. Whether any Hawking radiation reaches infinity depends on the relation between the emission domain and the event horizon. From the outside view, even if radiation is seen as always being emitted, the black hole never evaporates away, rather its
mass and entropy asymptote to finite non-zero limits, and the event horizon always acts as a sink for matter and information. From an inside view, the matter and information disappear into the singularity, which is the boundary of spacetime.  The argument is based on the nature of the processes at work plus a careful delineation of the relevant causal domains; in order to confirm this model and determine details of the outcome, detailed calculations of the expectation value of the stress-energy-momentum tensor are needed to determine back reaction effects.}
\end{abstract}

\section{Introduction}
\label{sec:intro}

This paper is about the final state of astrophysical black holes
forming from the collapse of a single astrophysical object. A current view \cite{HawPen96} with many proponents  is that in these circumstances, a singularity forms and is hidden by an event horizon, that is, a black hole comes into
being, but then the blackhole completely evaporates away within a
finite time due to Hawking radiation (\cite{Haw74},
\cite{BirDav84}). But there are other views. \\

\noindent According to Matt Visser (private communication), there are 3 basic scenarios for what will happen:
\begin{enumerate}
  \item complete evaporation with a regular endpoint,
  \item partial evaporation terminating in a stable remnant,
  \item partial evaporation terminating in a naked singularity.
\end{enumerate}
None of these options has been completely conclusively ruled out, however the first is the prevalent view. The conclusion of this paper is that it is the second that in fact occurs: the black holes that form as the final
state of collapsing stars are eternal rather than evaporating
completely away, as usually supposed. \\

As energy is radiated away, the surface $r=2m(u)$ where $u$ is a time coordinate along it and $m$ the interior mass, does not stay at the same $r-$coordinate value because $dm/dt<0;$ it moves
inwards, and consequently is not a null surface.  Thus it lies in the interior of the event horizon. It is a timelike marginally outer trapped 3-surface (to be called the EMOTS), a dynamical horizon \cite{AshKri02} that lies inside the event horizon and whose location is locally determined.\\

The key point of the argument is that Hawking radiation carries mass away from the EMOTS surface, rather than from the event horizon.  More precisely, Hawking radiation is emitted from an Emission Domain
that is locally determined \cite{Vis01} and lies just outside the EMOTS (\cite{ParWil00}, \cite{Cli08}).\\

Hawking radiation emitted from
the Emission Domain  at late collapse times ends up at the future
singularity inside the event horizon, rather than at infinity.
Indeed the existence of that future singularity is inevitable
because of the creation of closed trapped surfaces due to the
focusing of this outgoing radiation by Cosmic Microwave Background (CMB) radiation.\\

There is a second marginal outer trapped 3-surface (the OMOTS) that
is spacelike and lies outside the outgoing initial null surface generated by the initial marginally trapped
2-surface (the $I_{MOTS}$) that marks the onset of black hole dynamics. The OMOTS\
surface is also locally determined; it moves outwards with time because of incoming Cosmic Microwave Background radiation, and because of backreaction from Hawking radiation. Because it bounds the trapped domain in spacetime, it determines the outside edges of the future singularity. Thus it non-locally determines the location of the event horizon. It is a dynamical horizon whose properties characterise the mass and angular momentum of the interior body. \\

The final state of the black hole is a spacelike singularity that never
evaporates, situated in an asymptotically flat empty spacetime. It is
surrounded by an event horizon lying outside the OMOTS surface. The OMOTS surface is located at $r=2m_{final}$, given by
\begin{equation}\label{eq:final_m}
m_{final} = m_0 - \Delta m_{emit}
\end{equation}
where $m_0$ is the initial star mass, $\Delta m_{emit} \geq 0$ being the mass that succeeds in
escaping to infinity before the radiation emission surface disappears behind
the event horizon, and $m_{final}>0$. This suggests a Third Law of black hole
thermodynamics for uncharged black holes:\ it is not possible for radiation
processes to reduce the entropy of a black hole to zero. $\Delta m_{emit}$ may or
may not be non-zero; this depends on the relation between
the emission surface and the event horizon.\newline

The information loss paradox (\cite{Haw76}, \cite{Mat12}) is resolved by the resulting global spacetime
structure because, as seen from the exterior, a remnant mass $m_{final}$ given by (\ref{eq:final_m})
is left over at all times and does not evaporate away. As seen from the interior, microstates fall into the
future singularity and are lost there; they are no longer accessible to ordinary physics, because they have reached the boundary of spacetime. When that happens there is no longer an open domain within which local physical laws can be formulated. Unbounded tidal forces will also occur. This transition to a state of either non-existence, or at least being unable to interact, does not conserve energy (indeed energy is no longer defined) and is therefore clearly a non-unitary process. Of course in practice quantum gravity will come into play and determine what happens; but this will be happening in a domain that has no causal contact to the external universe, because it is behind the event horizon.\\

The paper considers these issues in the case of a single black hole with
spherical symmetry, where the relevant exterior solution is an exterior
Schwarzschild solution surrounding the single collapsing mass. The result is
probably stable in the case of more general geometries such as rotating
black holes, perturbed black holes, and if there are later infalling shells
of matter. The paper does not consider multiple black holes or black hole
collisions. Because it considers only astrophysically relevant situations,
it also does not consider charged black holes. \\

The argument is largely geometrical and heuristic, based on the idea of particle emission as presented for example by Parikh and Wilczek \cite{ParWil00}. Paul Davies, Malcolm Perry and Gary Gibbons have however emphasized to me that the idea of a particle is questionable in the context considered; there is no simple correlation between particles and the stress-energy-momentum in general, and the particles that an observer detects depend on the observer's world line as well as the quantum state. Additionally the notion of a particle is an observer dependent one. The thing that is objective is the energy momentum tensor which one might hope can incorporated into a self-consistent scheme a bit like Hartree-Fock. Paul Davies writes (private communication)
\begin{quote}
 ``\emph{The fate of the black hole depends on the back reaction
of the emission process, which will be described by the
semi-classical Einstein equation (or one's favourite variant
thereof) with a source term on the right being the expectation value
of the stress-energy-momentum tensor. (Issues arise about the
consistency of this semi-classical treatment, but most people feel
those issues become serious only as the Planck mass is
approached)}.''
\end{quote}
In summary, a particle view may be misleading; the only important quantity in calculating the back-reaction is the expectation value of the energy-momentum tensor. \\

Given these comments, perhaps much of the paper must be taken as  heuristic only, aimed at motivating work on the full quantum field theory calculations that can test and develop what is presented here. Nevertheless I believe what is presented may be a reasonable guide to the possibilities. It hinges on two key features: location of the Marginally Trapped surfaces; and location of an effective Emission Surface. The first is purely geometrical, and will be valid whenever we can expect a classical geometry to emerge. The second is certainly based in particle ideas, but because of the above comments, I have given proposed definitions of an Emission Domain and an Emission Surface, based in the energy momentum tensor. Thus these should be valid even when the particle idea is not. The issue then is where they lie, because that determines where the radiation ends up. \\

An outline of the paper is as follows.  Section 2 looks at classical black hole formation, and Section 3 at
the current canonical semi-classical picture. Section 4 proposes a
new semi-classical picture where the dynamical effects of mass loss
on trapping surfaces are properly taken into account, and Section 5
explores the causal nature of the eternal black holes that result.
Section 6 looks at the implications firstly for black hole
thermodynamics, proposing a new third law for uncharged black holes;
secondly for astrophysics; and thirdly for the information loss
paradox, which is resolved by this proposal because the black hole
never radiates totally away. Section 7 considers further work that
needs to be done to check and develop this proposal.

\section{Classical black hole formation}
\label{sec:classical}

In the case of classical general relativity, spherical gravitational
collapse of an astrophysical object of sufficient mass will lead to
formation of a MOTS 3-surface and consequently trapped surfaces and a black
hole will develop, characterised by existence of a singularity and an event
horizon that hides it from the outside world (\cite{Pen65},\cite{HawEll73},%
\cite{HawPen96}). The singularity is spacelike and the mass of the black hole
is unchanging:\ it is equal to the mass of the infalling star that created
it.

\subsection{Null geodesics and trapping surfaces}
\label{sec:null_geodesics}
The analysis depends on the propagation of radiation and nature of
trapping surfaces for null geodesics.

\subsubsection{Null geodesics}
Radiation propagates on irrotational null geodesics with affine parameter $%
\lambda $ and tangent vector $k^{a}(\lambda )$:
\begin{equation}
k^{a}=\frac{dx^{a}}{d\lambda },k_{a}k^{a}=0,k_{;b}^{a}k^{b}=0.
\label{eq:null_geod}
\end{equation}%
The divergence $\hat{\theta}$ of a bundle of these geodesics, given by $\hat{%
\theta}=k_{;a}^{a}$, determines how the cross sectional area $A(\lambda )$
of the bundle changes:\
\begin{equation}
\frac{1}{A}\frac{dA}{d\lambda }=\frac{1}{2}\hat{\theta}.
\label{eq:null_area}
\end{equation}%
The rate of change of $\hat{\theta}$ down the null \ geodesics is given by
the null Raychaudhuri equation (\cite{HawEll73}; \cite{HawPen96}:12):
\begin{equation}
\frac{d\hat{\theta}}{d\lambda }=-\hat{\theta}^{2}-2\hat{\sigma}_{ij}\hat{%
\sigma}^{ij}-R_{ab}k^{a}k^{b}  \label{eq:null_ray}
\end{equation}%
where $\hat{\sigma}^{ij}$ is the shear of the null geodesics, and the Ricci
tensor $R_{ab}$ is determined pointwise by the Einstein field equations.

\subsubsection{Trapping 2 surfaces}
The growing gravitational field of a collapsing star tends to hold light in,
hence decreases the divergence $\hat{\theta}_{+}$ of the outgoing null
geodesics in a spherically symmetric spacetime for any 2-sphere $%
S(r,t):=\left\{ r=const,t=const\right\} $ as $r$ decreases. The divergence $%
\hat{\theta}_{-}$ of the ingoing null geodesics is always negative;
the divergence $\hat{\theta}_{+}$ of the outgoing geodesics is
positive in flat space, as well as in a Schwarzschild solution for
$r>2m.$ A marginally trapped outer 2-surface $S_{MOTS}$ occurs when
the gravitational field due to the central mass is so large that
divergence $\hat{\theta}_{+}$ of the outgoing geodesics vanishes.

\begin{quote}
\textbf{Definition: Marginally Outer Trapped 2-Surface
$(S_{MOTS})$.} A spacelike 2-sphere is said to be a Marginally Outer
Trapped 2-Surface if the expansion $\theta _{+}$ of the outward null
normal vanishes:
\end{quote}
\begin{equation}  \label{eq:mots}
\hat{\theta}_{+}(S_{MOTS})=0.
\end{equation}%
This will happen in a Schwarzschild solution when $r_{_{MOTS}}=2m.$ For
smaller values of r, the 2-spheres $S(r,t)$ lying at coordinate values $r,t$
will be closed trapped surfaces $S_{CTS}$: that is,
\begin{equation}  \label{eq:cts}
r_{_{CTS}}<2m\Rightarrow \hat{\theta}_{+}(S_{CTS})<0.
\end{equation}%
Then if the energy condition%
\begin{equation}  \label{eq:null_energy}
R_{ab}k^{a}k^{b}\geq 0
\end{equation}%
is satisfied, by (\ref{eq:null_ray}) the outgoing null geodesics from $%
S_{CTS}$ will converge within a finite affine distance (\cite{HawPen96}:13)
and so will lie in the interior of the future of $S_{CTS}$ (\cite%
{HawPen96}:14). As these geodesics bound the causal future of the
2-sphere $S_{CTS}$ , this future will then be confined to a compact
spacetime region,
which implies a spacetime singularity must occur in the future of $S_{CTS}$ (%
\cite{Pen65}; \cite{HawEll73}; \cite{HawPen96}:28).

\subsubsection{Trapping 3-surfaces}\label{sec:dynamical}

We generalise the definition of dynamical horizons by Ashtekar and
Krishnan \cite{AshKri02} by removing the restriction that it be a
spacelike surface. Thus,
\begin{quote}
\textbf{Definition: Dynamical Horizon.} A  smooth, three-dimensional
sub-manifold $H$
in a space-time is said to be a \textit{%
dynamical horizon }if it is foliated by a preferred family of
2-spheres such that, on each leaf $S$, the expansion $\theta _{(\ell
)}$ of one null normal $\ell _{a}$ vanishes and the expansion
$\theta _{(n)}$ of the other null normal $n_{a}$ is strictly
negative.
\end{quote}
 Consequently a dynamical horizon $H$ is a
3-manifold which is foliated by marginally trapped 2-spheres. On this definition, such 3-surfaces can be timelike, spacelike, or null, with rather different properties. The essential point is that
they are locally defined, and therefore are able respond to local
dynamic change. You don't need to know what is happening at infinity
in order to determine a local physical effect.\\

Dynamical horizons are \emph{Marginally Outer Trapped 3-Surfaces}
(MOTS). The expansion $\theta _{(\ell )}$ of the outgoing null
normal  $\ell _{a}$ vanishes on the MOTS, and is either positive in
the outside region, when it will be called an OMOTS, or negative,
when it will be called an EMOTS. The two different kinds of MOTS
surfaces have crucially different properties. An OMOTS is the outer
bound of a
trapping domain; an EMOTS is the inner bound. This difference can be expressed in terms of derivatives of the
divergence $\hat{\theta}$ of the outgoing null geodesics. As in
\cite{AshKri02}, consider outgoing null geodesics to the 2-spheres
$S(v,r)$ with tangent vector $\ell^a$, and ingoing null geodesics
with tangent vector $n^a$. Then the MOTS\ 3-surfaces, defined by
$\theta _{(\ell )}=0$, are associated with the gradient of $\theta
_{(\ell )}$ in the $n^a$ direction, which determines if the MOTS
3-surface is timelike or spacelike  as follows \cite{AshKri02} (for
a proof, see  \cite{Dreetal03}: Section IIB).
\\

\begin{center}
$%
\begin{array}{|c|l|l|l|l||}
\hline \hline \text{MOTS 3-surface}  &
\partial \theta _{(\ell )}/\partial n^{a}: & \text{Nature:} & \text{Trapped region}:\\ \hline \hline \text{EMOTS}  &  \partial \theta
_{(\ell )}/\partial n^{a}>0 & \text{timelike} & \text{Inner bound} \\
\hline \text{OMOTS}   &
\partial \theta _{(\ell )}/\partial n^{a}<0 & \text{spacelike}  & \text{Outer bound}
\\ \hline \hline
\end{array}%
$
\newline
\end{center}

\textbf{Table 1:} \emph{Relation between trapping properties and
causal character of non-null MOTS 3-surfaces}.\\

In the case of a null MOTS, $\partial \theta _{(\ell )}/\partial n^{a}$ can have either sign, as can be seen from the maximally extended Schwarzschild solution, or indeed can vanish, as can be seen from plane wave solutions. Thus a null MOTS can be either an EMOTS or an OMOTS.

\subsection{Eternal Black Holes: The Schwarzschild solution}
\label{sec:Schw} Eternal black holes are described by the maximal
extension of the Schwarzschild spherically symmetric vacuum
solution, characterised by its mass $m$.

\subsubsection{Coordinates and metric}
The exterior part ($r>2m)$ is given in standard spherical coordinates by \cite{HawEll73}
\begin{equation}
ds^{2}=-(1-\frac{2m}{r})dt^{2}+\frac{dr^{2}}{(1-\frac{2m}{r})}+r^{2}(d\theta
^{2}+\sin ^{2}\theta d\phi ^{2}) .  \label{eq:Sch}
\end{equation}%
Proper time along a timelike world line $\{r=\text{const},\theta =\text{const%
},\phi =\text{const}\}$ for $r>2m$ is given by
\begin{equation}
\tau =\int \sqrt{(1-\frac{2m}{r})}dt .  \label{eq:propertime}
\end{equation}%
Incoming radial null geodesics obey
\begin{equation}  \label{eq:incoming_null_geod}
\left\{ ds^{2}=0,d\theta =d\phi =0\right\} \Rightarrow \int dt=\int \frac{dr%
}{(1-\frac{2m}{r})}.
\end{equation}%
With (\ref{eq:propertime}), this immediately gives the gravitational
blueshift $z$ for incoming radiation from infinity:
\begin{equation}
\frac{\nu(r)}{\nu _{E}}=\sqrt{ {\frac{1-2m/r_{E}}{1-2m/r}}} \rightarrow
\frac{\nu(r)}{\nu _{\infty }}=\sqrt{ {\frac{1}{1-2m/r}}}
\label{eq:blueshift}
\end{equation}%
as the radius of emission $r_{E}\rightarrow \infty.$

\subsubsection{Extension to $r<2m$}
The metric (\ref{eq:Sch})\ is singular at $r=2m$, but there exist regular
coordinates across this null surface \cite{HawEll73}. Define
\begin{equation}  \label{eq:Schw_rstar}
r^{\ast }=\int \frac{dr}{1-2m/r}=r+2m\log (r-2m);
\end{equation}%
then $v:=t-r^{\ast }$is an advanced null coordinate and (\ref{eq:Sch}) can
be rewritten in the Eddington-Finkelstein form in terms of the coordinates $%
(v,r,\theta ,\phi )$ as
\begin{equation}  \label{eq:Edd-Fink}
ds^{2}=-(1-\frac{2m}{r})dv^{2}+2dvdr+\ r^{2}(d\theta ^{2}+\sin ^{2}\theta
d\phi ^{2})
\end{equation}%
which is regular for all $r>0.$ The solution is static for $r>2m$
(as $\partial/\partial t$ is a timelike Killing vector field), and
hence that region is eternal (it does not come to an end at any
finite positive or negative value of the time parameter $t$), but is
dynamic for $r<2m$. The maximal extension employs several such
coordinate patches \cite{HawEll73}, but that need not concern us
here, as we will focus on astrophysical black holes that form by
spherical  objects collapsing in the context of a Schwarzschild
vacuum solution.

\subsubsection{Trapping surfaces}
Each 2-sphere $S(v,r)$ satisfying $r=2m$ is a $S_{MOTS}$, for all
values $v$. Its ingoing normal null geodesic family is generated by
the null vectors $l^{a}=\frac{dx^{a}}{d\lambda }=(0,1,0,0)$ so these
geodesics are given by $x^{a}=(v,\lambda,\theta _{0},\phi _{0})$ and
link the 2-spheres $S(v,\lambda )$ to each other ($v$ labels the
2-spheres in the OMOTS 3-surface $r=2m$). Its outgoing normal null
geodesic family is generated by the vectors
$k^{a}=\frac{dx^{a}}{d\lambda }=(1,0,0,0)$ which
are null for $r = 2m$, so these geodesics are given by $x^{a}=(\lambda,2m,%
\theta _{0},\phi _{0})$ and link the $S_{MOTS}$ 2-spheres $S(\lambda,2m)$ to each
other. The metric of these 2-spheres is $ds_{S(\lambda,2m)}^{2}=(2m)^{2}(d%
\theta ^{2}+\sin ^{2}\theta d\phi ^{2}),$ with constant area $16\pi m^{2}$
for all $\lambda.$ This demonstrates that the outwards geodesics are
non-diverging:$\ \hat{\theta}_{+}(S(v,2m))=0.$ This area is constant for all
$\lambda $ because equation (\ref{eq:null_ray})\ is satisfied with $\hat{%
\theta}=0,\ \hat{\sigma}_{ij}\hat{\sigma}^{ij}=0,R_{ab}k^{a}k^{b}=0.$ These
geodesics generate the event horizon given by
\begin{equation}
r_{EH}=2m,\;m=const.  \label{eq:rh}
\end{equation}%
because events for $r<r_{EH}$ cannot send signals to the outside domain $%
r>2m.$ \newline

In summary:\ the unique family of $S_{MOTS}$ 2-surfaces in the
Schwarzschild solution occur at $r=2m$ for all $v,$ and constitute
the event horizon, which is also an OMOTS surface. Conversely, each 2-sphere $S(v,2m)$ comprising the
event horizon is a $S_{MOTS}$ 2-surface. This equivalence will no
longer be true when Hawking radiation occurs and we take
backreaction into account in the collapsing star case, as discussed below.

\subsection{Astrophysical black hole formation} \label{sec:class_bh}
When astrophysical black holes form, the exterior vacuum Schwarzschild solution is joined
on to an interior collapsing fluid solution. The exterior part of the solution is static, because of Birkhoff's theorem \cite{HawEll73},
even though the fluid part - the source of space-time curvature - is dynamic. Mass is conserved by the dynamics.

\subsubsection{Domains}
 To model such classical black hole formation by collapse of a
spherically symmetric star, consider three domains: a vacuum region
outside the infalling fluid and outside the trapped region; a
trapped vacuum region outside the infalling fluid; and the infalling
fluid. This is shown in the
conformal Penrose diagram of the collapse (\cite{HawEll73}, \cite{HawPen96}%
:42,63), see Figure 1, where every point represents a 2-sphere $S(t,r)$.
Thus, we have
\begin{figure}[tbp]
\includegraphics[width=7in]{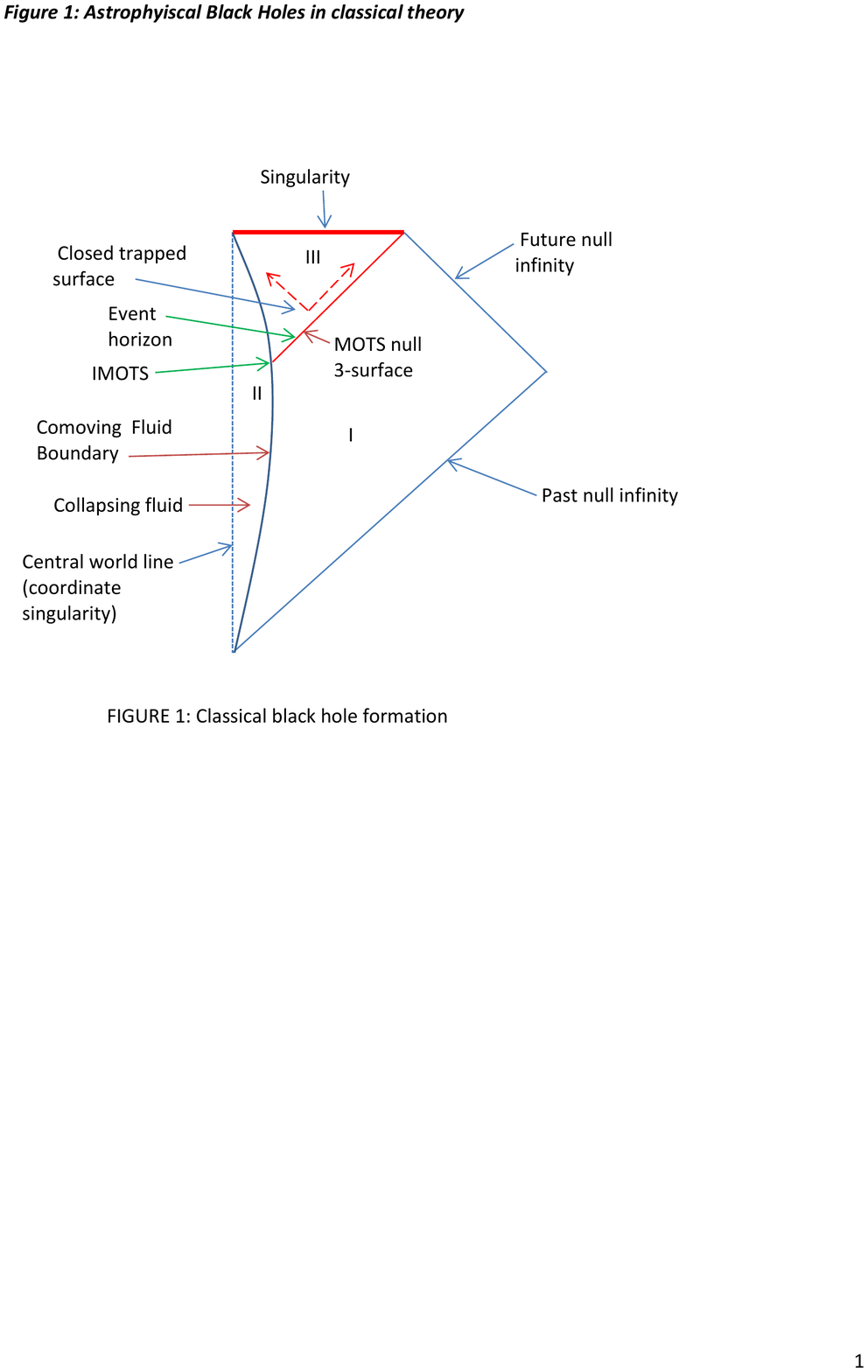}
\caption{Figure 1}
\label{Fig1}
\end{figure}

\begin{enumerate}
\item \textbf{Domain I:} An outer vacuum spacetime. bounded on the outer sides by past and future null infinity,
and on the inner side by the fluid boundary and the event horizon. In the spherically
symmetric case, which is what we consider in this paper, it is a
Schwarzschild solution. It is static.

\item \textbf{Domain II}: The infalling fluid: pressure free matter or a
perfect fluid obeying the usual energy conditions. In simple cases this can
be modeled as a portion of a spatially homogeneous\ Friedmann-Lema\^{\i}tre
model. It is bounded on the outer side by the vacuum Schwarzschild solution.

\item \textbf{Domain III}:\ The vacuum spacetime outside the fluid sphere
but inside the event horizon. It is a spatially homogeneous but time
evolving part of a Schwarzschild solution. It is comprised of closed trapped
surfaces, which imply existence of the future spatial singularity that
bounds Domain III to the future.
\end{enumerate}

\subsubsection{Boundaries}
The boundaries between the domains are,
\begin{itemize}
\item \textbf{B12:} Between I and II, the comoving Fluid Boundary with radius $r =
R_S(\tau)$,

\item \textbf{B23:} Between II and III, the comoving Fluid Boundary $r = R_S(\tau)$,

\item \textbf{B13:} Between I and III, the event horizon given by (\ref{eq:rh}), which is
a Killing horizon and the locus of a family of $S_{MOTS}$ 2-spheres, i.e.
each 2-sphere in the event horizon is a $S_{MOTS}$ with
$\hat{\theta}_+=0$. Objects inside $r_{H}$ are trapped, because
$r=r_{H}$ is a null surface. The event horizon hides the interior
from the exterior, just as it does in the case of a maximal
Schwarzschild solution.
\end{itemize}
The inmost $S_{MOTS}$ 2-surface (IMOTS) is the 2-sphere at the join
of domains I, II, and III; in terms of the collapse of the fluid,
this is the event marking the start of the enclosure of the fluid by
closed trapped 2-surfaces.

\subsubsection{Trapping surfaces}
The nature of the trapping surfaces in this solution are as follows
(see Figure 2, where one can easily see what is the sign of the change of $\hat{\theta}_+$ as one moves to the left over each MOTS surface):
\begin{figure}[tbp]
\includegraphics[width=7in]{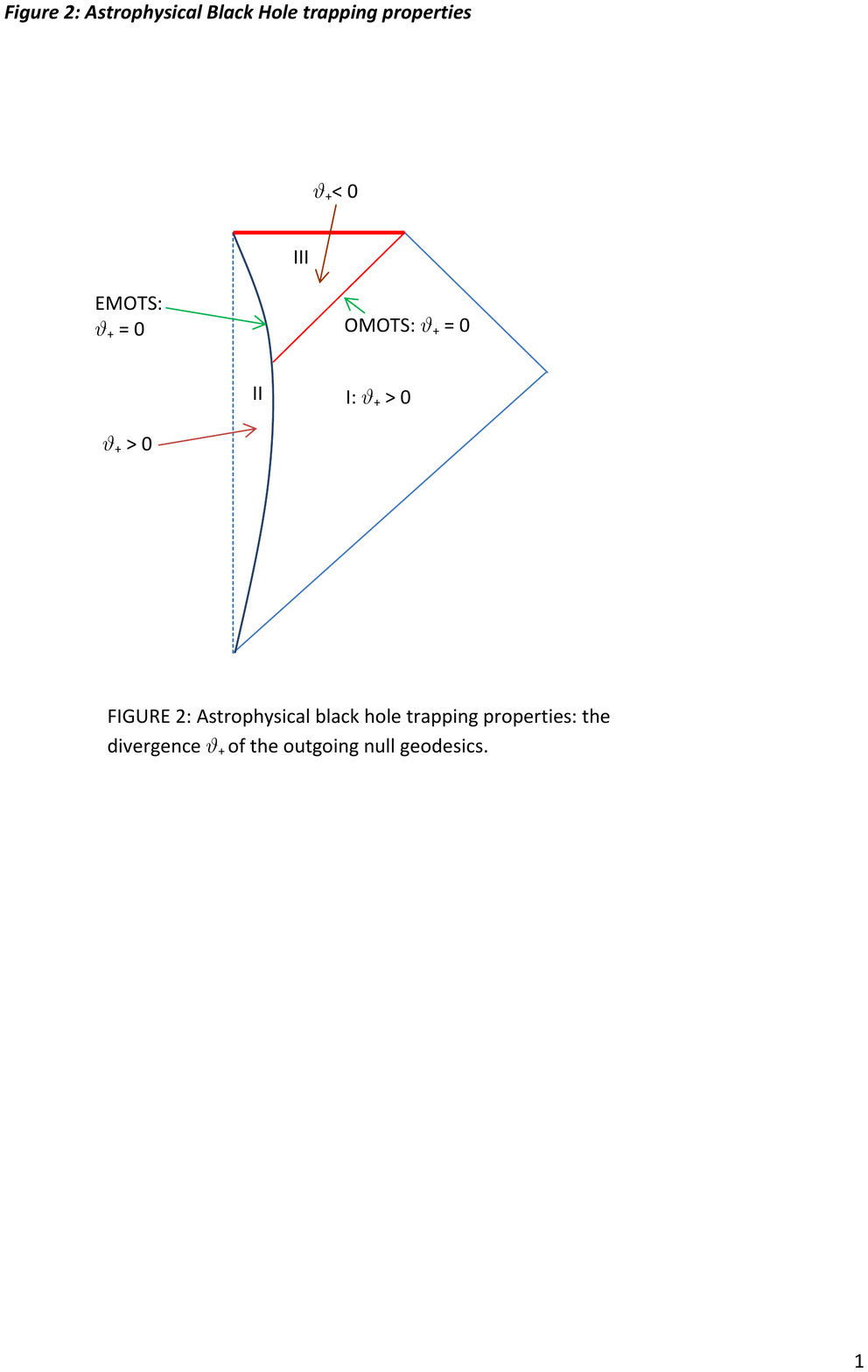}
\caption{Figure 2} \label{Fig2}
\end{figure}
\begin{itemize}
  \item The event horizon $r=2m$ in the Schwarzschild vacuum is a null OMOTS separating Domain I and Domain III.
 \item the fluid surface boundary between domains II and III is a
  timelike EMOTS.
  \item These two MOTS 3-surfaces intersect in the IMOTS 2-surface.
  Hence there is a bifurcate trapping surface structure in this case.
\end{itemize}
If one considers surface of constant coordinate time in the obvious
sense (they are horizontal lines in this diagram), for small values
of $t$ they encounter only outward null diverging (OND) 2-spheres in
domains I and II, so there are no MOTS surfaces. When at time
$t_{IMOTS}$ the constant time surfaces intersect the IMOTS 2-sphere defined by
where the null surface $r=2m$ meets the fluid surface, two
counterpoising MOTS 3-surface appear: the OMOTS spacelike surface
heading out, and the timelike EMOTS surface heading in. For times
greater than $t_{IMOTS}$, the constant time surface intersect the inner OND
region in the fluid interior (Domain II); the timelike EMOTS
surface; vacuum Domain III which is filled with outward null
converging (ONC) 2-spheres; the spacelike OMOTS surface; and the
outer OND vacuum. Hence as $t$ increases through $t_{IMOTS}$ there
is a bifurcation from no MOTS surfaces to 2 oppositely signed MOTS
surfaces.

\subsection{Outcomes}
A Ricci tensor singularity will occur inside the fluid as the fluid
collapses and $R_S(\tau) \rightarrow 0$ within a finite proper time.
This may or may not be accompanied by a Weyl singularity inside the
fluid; such a singularity will not occur for example if the fluid is
a Friedmann-Lema\^{\i}tre model, as the Weyl tensor is then zero
inside the fluid. The Ricci tensor is zero outside the collapsing
star; it is the Weyl tensor that diverges at the spacelike
Schwarzschild singularity in the vacuum domain. Specifically, the
Kretschman scalar is
\begin{equation}  \label{eq:Kretsch}
K\ =C_{abcd}C^{abcd}=\alpha \frac{m^{2}}{r^{6}}
\end{equation}%
where $\alpha =48G^{2}/c^{4}$ \cite{Hen99}, so this diverges as $%
r\rightarrow 0.$ It is the spatial inhomogeneity of the matter distribution
(non-zero density inside the fluid, zero outside)\ that generates this
singularity in the conformal structure of spacetime.\\

As seen from the outside, the mass of the star never alters; it is always equal to the initial value $m_0$:
\begin{equation}\label{eq:mass_const}
  m = m_0 = \textit{const}.
\end{equation}
This will of course not be the same if matter falls into the blackhole, hereby increasing it is mass; then the horizon
is a dynamic horizon \cite{AshKri02} and the laws of black hole thermodynamics \cite{BarCarHaw73} come into
play to characterise the resulting changes.
However we do not consider those processes here.

\subsection{The Effect of Cosmic Background Radiation}\label{sec:cbr}
Cosmic microwave blackbody background radiation pervades the
universe. It alters the causal domains significantly.

\subsubsection{Cosmic Background Radiation}\label{sec:cbr1} \
The universe is permeated by cosmic microwave blackbody (CMB)
radiation \ emitted from the hot big bang era in the early universe
(\cite{PetUza09}, \cite{EllMaaMac12}). That blackbody background
radiation was emitted by the last scattering surface at the end of
the Hot Big Bang era in the early universe, it then propagates
through the universe with a temperature
\begin{equation}
T_{CMB}=(1+z)\,T_{CMB}|_{0}  \label{eq:CMB_Temp}
\end{equation}%
where $z$ is the redshift characterising the cosmological time when
the radiation temperature is measured. It now pervades the universe
at a temperature
\begin{equation}
T_{CMB}|_{0}=2.7K.  \label{eq:cmb_0}
\end{equation}%
In effect, this radiation reaches local systems at the present time
from an effective \textquotedblleft Finite
Infinity\textquotedblright (\cite{Ell84}, \cite{Ell02}) at about one
light year's distance that emits blackbody radiation at a
temperature $T_{CMB}|_{0}.$ This is the effective sky for every
local system. The CMB radiation has the stress tensor of a perfect fluid with
positive energy density and pressure:
\begin{equation}
p_{CMB} = \rho_{CMB}/3,\,\, \rho _{CMB}=a\, T_{CMB}^{4}.
\label{eq:cmb_density}
\end{equation}
It has an entropy density given by
\begin{equation}
S_{CMB} = \frac{4}{3} a T_{CMB}^3 +b \label{eq:cmb_entropydensity}
\end{equation}
where $b$ is a constant independent of the energy $E$ and volume $V$
(\cite{Lem13}:25).

\subsubsection{A Spacelike OMOTS inside the event horizon }\label{sec:OMOTS_revise} \
In particular, this radiation will be the environment for black
holes, and will permeate Domain I and hence will cross the even
horizon and permeate Domain III also. In doing so, the effect of the
CMB is to displace the OMOTS trapping surface, thus creating a new causal domain (see Figure 3).\\
\begin{figure}[tbp]
\includegraphics[width=7in]{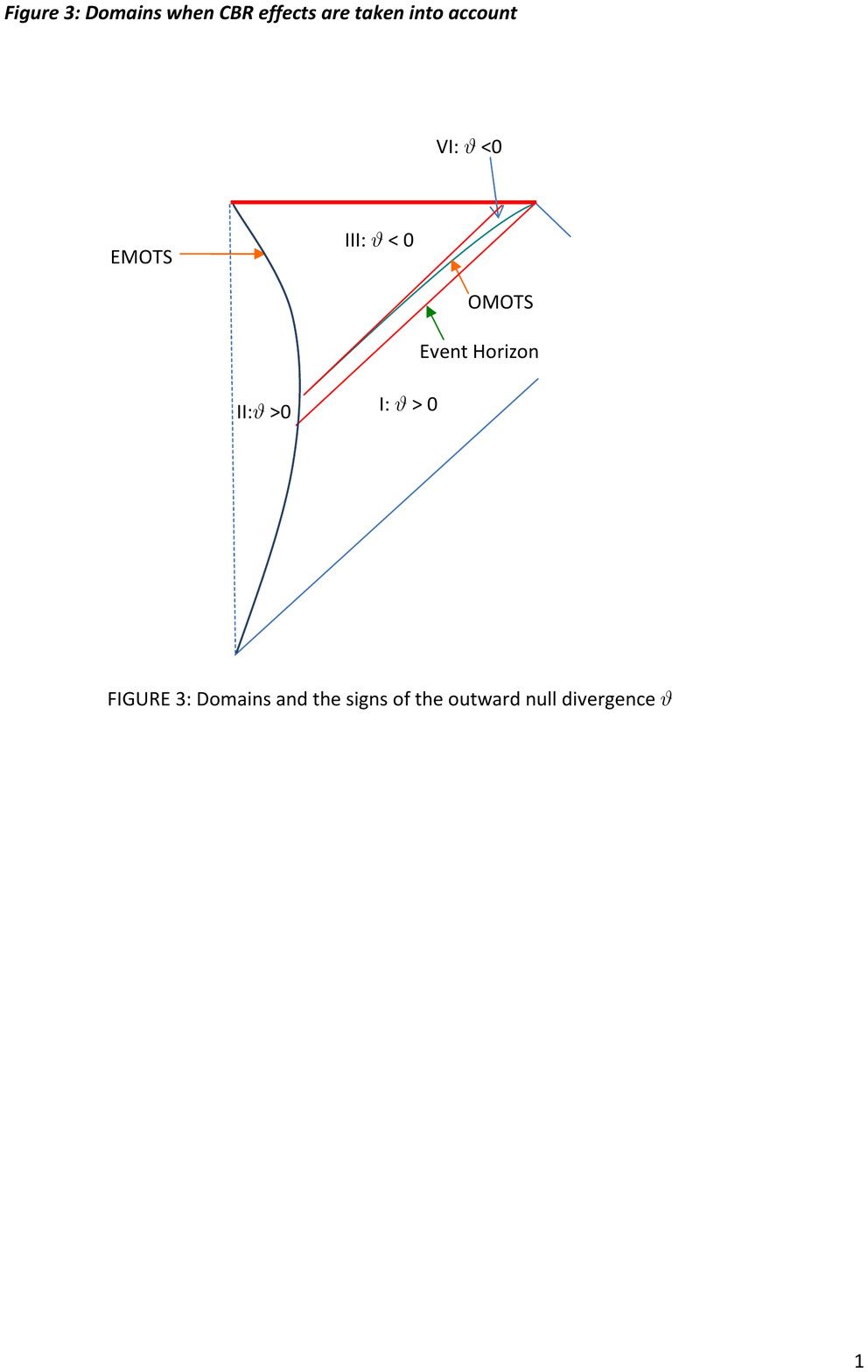}
\caption{Figure 3} \label{Fig3}
\end{figure}

Because of the null Raychaudhuri equation (\ref{eq:null_ray}), the
incoming CMB with density and pressure (\ref{eq:cmb_density}) causes
focussing of null geodesics, so any outgoing null geodesics that
start at an initial affine parameter $\lambda_1$ at a $S_{MOTS}$
2-surface where $\hat{\theta}_+ = 0$ will converge to a conjugate
point at finite affine parameter value $\lambda_2\,>\,\lambda_1$.
Hence a marginally trapped $S_{MOTS}(\lambda_1)$ in the OMOTS
3-surface will be mapped by the outgoing null geodesics to trapped
2-spheres $S_{CTS}(\lambda)$ with $\hat{\theta}(\lambda) < 0$ for
$\lambda_1\,<\,\lambda\,<\,\lambda_2$. These 2-spheres, whose future
necessarily ends on a singularity (because they are trapped!), will
therefore lie inside the OMOTS surface; hence the OMOTS surface has
moved out relative to the null geodesics through
$S_{MOTS}(\lambda_1)$. \\

Consequently, the effect of the infalling CMB is to change the OMOTS
3-surface from null to spacelike. This is in accord with the
discussion by Dreyer et al \cite{Dreetal03}, who state that
infalling matter or radiation causes an isolated horizon to be
spacelike.  It now lies outside the initial null surface $r=2m$
generated by the geodesics through the IMOTS 2-sphere. It bounds the
domain of trapped surfaces, and so it bounds the events whose future
ends up at the singularity. Thus the OMOTS in fact defines the
extent of the future singularity (in the conformal diagram Figure
3). We denote the point where it
reaches the future singularity as $P_2$; this is the outer edge of the singularity.\\

The further important consequence is that the event horizon is no
longer where it used to be. It used to be generated by the outgoing
null geodesics starting at the IMOTS 2-sphere where $r=2m$, and reaching the
future singularity at the point $P_1$. That initial null surface is
now trapped. The question is which domain in this diagram is
trapped, that is, future timelike and null geodesics from points in
this domain necessarily end up on the future. The answer is that it
is the domain bounded by the null geodesics through $P_2$; these in
fact locate the event horizon, which lies outside the IMOTS. Unlike the MOTS surfaces, the event horizon does not bound any locally significant domains, and so cannot be found from any local property. To determine its position you need to know about all future matter and radiation that may fall in across the OMOTS surface from the IMOTS 2-sphere until infinity.  \\

Now one might think this effect is only a matter of principle and
will be completely negligible for example in the case of the black
hole at the centre of the Milky Way galaxy. However this is not
necessarily so, because of the unbounded blueshift the CMB photons
experience as they fall into the event horizon (let $r\rightarrow2m$
in (\ref{eq:blueshift})). This divergence means that the photon
energies diverge (cf. the discussion in Section
\ref{sec:revised_focussing}). Furthermore, this divergence in the density
will remain true no matter how small the incoming radiation density
is, provided it is non-zero; and as seen from outside the radiation
will continue to be present forever, albeit decaying away. Remember
that the size of the event horizon is not determined by the mass or
size of an object today: that just gives its minimum size. Its
actual size today has to do with how much energy crosses the OMOTS
surface from the present day to the end of time; and there is no
discernible marker present showing where it is today. Because it's location is determined by an integrated effect over an unbounded interval, the cumulative result may not be negligible. In
any case the fact the OMOTS surface is spacelike is conceptually
crucial, both because it shows this surface lies inside the event
horizon, and because it shows the analysis of Ashtekar and Krishnan
\cite{AshKri02} can be applied to this surface, allowing
determination of angular momentum, energy fluxes, and area balances. \\

\textbf{In summary:} the effect of the CMB is to change the OMOTS
3-surface from being a null surface that coincides with the event horizon,
to being a spacelike surface that
\begin{enumerate}
  \item lies outside the initial null surface,
  \item determines the extent of the future spacelike singularity,
and
  \item thus determines the location of the event horizon.
\end{enumerate}

\subsubsection{Revised domains}\label{sec:domains_revise} \
Consequent on this, there is a new causal domain that needs to be
recognized:
\begin{itemize}
\item \textbf{Domain VI}, a trapped region between the initial null surface
and the spacelike OMOTS surface, which lies outside the Initial Null
Surface.
\end{itemize}
As to the boundaries of this domain, they are
\begin{itemize}
  \item \textbf{B16}: Between I and VI, the
spacelike OMOTS surface that ends up at the limit point $P_2$ of the
spacelike singularity;
  \item \textbf{B36}: Between III and VI, the initial null surface through the IMOTS 2-sphere that ends up at
the point $P_1$ on the spacelike singularity.
\end{itemize}
One can note that the event horizon is not a boundary between any of
these domains.

\section{Semiclassical gravity:\ the standard view}
\label{sec:semiclassical}
This is all altered when one takes quantum field theory into account, so that Hawking radiation results in mass loss and (\ref{eq:mass_const}) is no longer true.
\subsection{The basic view}
\label{sec:semiclass_basic}

Collapse and horizon formation proceeds as before. Three new effects are
believed to occur.\newline

\textbf{Item 1}: Quantum field theory fluctuations leads to production of
black body Hawking radiation at the event horizon (\cite{Haw73}; \cite%
{BirDav84}, \cite{HawPen96}:43), with temperature determined by the mass of
the central body:
\begin{equation}
T_{BH}=\frac{1}{8\pi m}.  \label{Eq:T}
\end{equation}
This result has been calculated in many ways, and is independent of the
gravitational field equations \cite{Vis01}. The basic process can be thought
of as follows: virtual pair production takes place all the time in the
quantum vacuum. If one of a pair of such particles created near the black
hole falls behind the event horizon, it gets separated from the other, and
they become a real pair of particles, one falling into the black hole and
one being radiated outwards. Thus photons are continually emitted by the
black hole. Consequently in thermodynamic terms, the black hole acts as a
black body with temperature $T_{BH}$ given by (\ref{Eq:T}) and entropy
\begin{equation}  \label{Eq:S}
S_{BH}=4\pi m^{2}=\frac{1}{4}A.
\end{equation}
This latter result is dependent on the gravitational field equations \cite%
{Vis01}. The equation of state that follows from this is
\begin{equation}\label{eq:eos}
  S_{BH} =   \frac{1}{16 \pi T_{BH}^{2}}.
\end{equation}
\\

\textbf{Item 2}: Because radiation is being emitted, energy conservation
shows there must be a corresponding mass loss by the black hole:
\begin{equation}
dm/d\tau <0,  \label{eq:dmbydu}
\end{equation}%
where $\tau $ is give by (\ref{eq:propertime}) in the outer domain, so $m$
decreases with $\tau $ and therefore $T_{BH}$\ increases as the radiation
process continues. One can think of the outcome as like a Schwarzschild
solution with ever decreasing mass $m(\tau )$.\newline

\textbf{Item 3}: The singularity that has formed at the centre eventually pops
out of existence \cite{Haw74}, because the Hawking process inevitably
carries all the mass away to infinity. This happens in a finite time,
because as the mass decreases the radiation loss process speeds up. The
power $P$ radiated is proportional to $1/m^{2}$; using the usual mass-energy
equivalence,
\begin{equation}
P=-c^{2}dm/d\tau ,  \label{eq:power}
\end{equation}%
the evaporation time $t_{evap}$ is proportional to $m_{0}^{3}$ (\cite%
{Haw76}, \cite{Pag76}). Putting in the numbers, this lifetime will be of the
order
\begin{equation}
t_{evap}\simeq 10^{71}(m/m_{\odot })^{3}secs.  \label{eq:lifetime}
\end{equation}

\subsection{Causal Structure}
\label{sec:semiclass_domains}
The related causal structure is characterised by the relevant domains and their boundaries.

\subsubsection{Domains}

Three of the spacetime domains in this case are similar to the
previous case, but the Penrose diagram is modified by addition of a
new domain at late times (\cite{BirDav84}: 272; \cite{HawPen96}:63;
\cite{LidBro07}) (Figure 4).
\begin{figure}[tbp]
\includegraphics[width=7in]{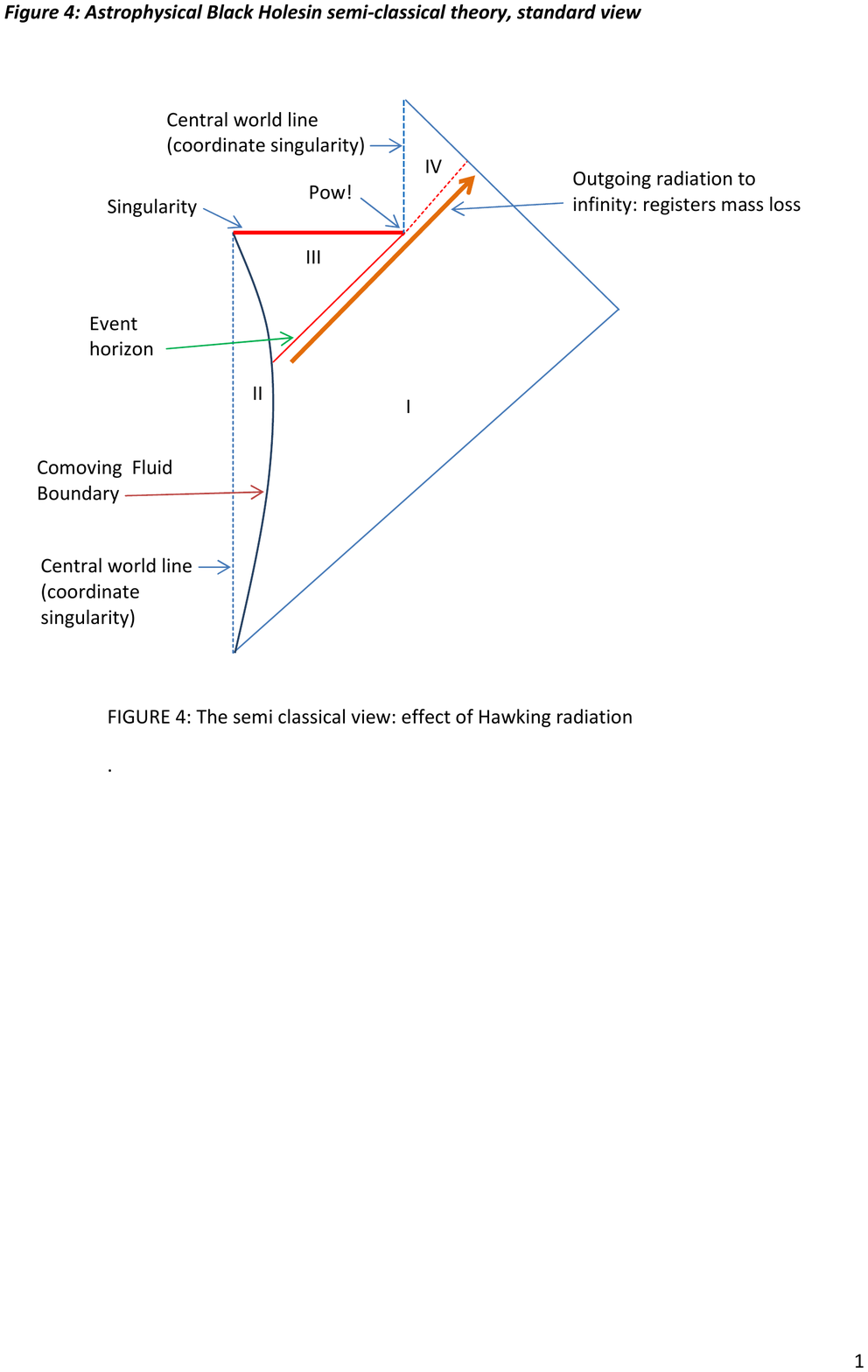}
\caption{Figure 4} \label{Fig4}
\end{figure}
The domains are,

\begin{enumerate}
\item \textbf{Domain I}, an exterior Schwarzschild solution modified by
having varying mass and outgoing radiation, that can perhaps be modelled by a Vaidya outgoing
solution (\cite{Vaidya1}, \cite{Vaidya2}). It is bounded on the outer sides by past and future null infinity. It is bounded on the inner side by the
event horizon $r=2m_{0},$ where $m_{0}$ is the initial mass of the infalling
object. Outgoing radiation to infinity registers the mass loss at the centre.

\item \textbf{Domain II}, infalling matter as before. Again it can for
example be a flat FRW dust model $(0<r<R_{S}(\tau ))$ collapsing within
domain I.

\item \textbf{Domain III,} the domain inside the event horizon and outside
the fluid:\ a Schwarzschild solution but with varying mass and ingoing
radiation, that is an ingoing Vaidya solution bounded on the inner side by
the fluid.

\item \textbf{Domain IV}:\ Minkowski spacetime, which must exist at late
times because the mass has evaporated to zero.
\end{enumerate}
\subsubsection{Boundaries}
The separation surfaces are as before:

\begin{itemize}
\item \textbf{B12:} Between I and II,  the fluid boundary up to the
2-sphere IMOTS where the event horizon forms

\item \textbf{B13:} Between I and III, the event horizon $r=2m_{0}$.

\item \textbf{B23:} Between  II and III, the fluid boundary after the IMOTS.

\item \textbf{B14:} Between I and II, a null surface.
\end{itemize}

\subsection{Outcomes}
The crucial new feature is that the
effect of the Hawking radiation is to eventually make both the
central mass and singularity vanish. How this happens is not very
clear, but the result is that the outer domain eventually loses the
event horizon and again has a regular centre.\newline

A key feature is that because the mass is decreasing as a function of time,
the MOTS surface is now time dependent: $r=2m(v)$ in terms of a suitable
time parameter $v$. This surface is therefore no longer the same as the
event horizon:\ rather it lies inside the event horizon. However it is
supposed this feature makes no difference to the overall outcome: it does
not affect the causal domains indicated above. This is the feature that will
be challenged below.

\subsection{Problems with the standard view}
\label{sec:semiclass_Prob}

While the physical motivation is clear - once Hawking radiation has
initiated, there seems to be nothing to stop it until the black hole has
radiated fully away - the associated Penrose diagram (\cite{HawPen96}:63;
\cite{LidBro07}) is rather puzzling \cite{VasStoKra07} and it is unclear
what happens (\cite{BirDav84}: 273).

\begin{itemize}
\item How does the external domain know the mass is supposed to have decreased? The
mass that is decreasing is far inside the event horizon; how does that
information get to infinity?

\item How can the radiation increase as the mass decreases if it is emitted
at the event horizon $r=2m_{0}$, which does not know about the mass loss in
the interior?\ The event horizon stays at $r=2m_{0}$ as the interior mass
decreases:\ if it were the source of radiation, the temperature would not
increase with time as the black hole\ evaporates.

\item How can the spacelike singularity go away once it has formed? We have
no handle on its dynamics: it is after all singular, and there is no known
set of equations to govern its dynamics.\

\item The basic argument used in detailed calculations that gives this
result (e.g. \cite{Haw76}) is a quantum field theory argument. Is it
legitimate to make such a prediction about strong quantum gravity effects on
this basis? A full quantum theory of gravitation will presumably come into
operation: it might well give a completely different outcome (indeed if it
does not, then why do we need the full quantum gravity theory?)
\end{itemize}

\section{A revised proposal}
\label{sec:revised}

To meet these problems, one can make a revised proposal that fully allows
for the mass of the central object changing with time in a consistent way.\\

This leads to a revised version of the relevant spacetime domains. This
section looks at the processes involved. The following sections look at the resulting
global structure, leading to the conclusion (Section \ref{sec:eternal}) that
we can expect there to be an eternal black hole.

\subsection{The overall view}
\label{sec:revised_overall}
The basic view is as follows: assume the gravitational field of a collapsing object becomes so high that closed trapped surfaces eventually occur around the collapsing fluid. An innermost trapped 2-sphere (the $I_{MOTS}$)  will occur on the surface of the fluid, marking the start of the existence of the black hole. \\

The initial null surface that spreads out from the $I_{MOTS}$ would be the event horizon if it were not for Hawking radiation on the one hand and incoming CMB radiation on the other. These result in a spacelike marginally outer trapped 3-surface (an OMOTS) existing outside the initial null surface, the OMOTS being the outer boundary of the domain  where trapped 2-spheres occur. This surface is spacelike and keeps growing because of infalling CMB radiation (Section \ref{sec:cbr}). A spacelike singularity will occur in the future of this surface because the Cosmic Microwave Background radiation (CMB) will make it's outgoing null geodesics converge. For this reason it determines the location of the event horizon (cf. Figure 3). The geometry of such dynamic horizons is very nicely discussed by Ashtekar and Krishnan \cite{AshKri02} and by Dreyer et al \cite{Dreetal03}. They show that this 3-surfaces is necessarily space-like for accreting black holes. One can determine the mass and angular momentum of the central object from the properties of such surfaces.  In summary:
 \begin{itemize}
 \item \textit{Does an event horizon occur}? The answer is yes; and its location is determined by the OMOTS surface.
 \end{itemize}

However there is also another marginally outer trapped 3-surface that arises at the IMOTS 2-sphere. It is a timelike EMOTS just outside the fluid sphere, being the inner boundary of the domain where trapped 2-spheres occur (cf Figure 3). Quantum field theory then shows that Hawking radiation will be emitted just outside the  EMOTS surface because of its local properties (\cite{ParWil00}, \cite{Vis01}, \cite{Cli08}.) Mass loss takes place as given by the standard formula (\ref{eq:power}), hence the mass is a decreasing function of time and the Schwarzschild radius decreases:%
\begin{equation}
r_{H}=2m(v),\;dm/dv<0,  \label{eq:rh2}
\end{equation}%
where $v$ is a time parameter along the developing local horizon.
The 2-surfaces comprising this EMOTS surface, defined as the inmost set of 2-spheres
where $\hat{\theta}_{+}=0$, will to start with instantaneously be
almost the same as in the corresponding Schwarzschild solutions with
time varying mass: $r_{EMOTS}=r_{H}(v).$ The 3-surface they generate
therefore moves inwards as mass is radiated away. It is a timelike surface comprised of marginally outer trapped 2-surfaces at each instant. Thus it is a timelike dynamic horizon.%
As this surface is timelike, we can choose $v$ as proper time along the
inner EMOTS\ 3-surface bounding the fluid sphere. The temperature of the
emitted radiation
\begin{equation}
T=\frac{1}{8\pi m(v)}  \label{eq:Mu2}
\end{equation}%
gets larger and larger because the radiation is emitted close to the
EMOTS\ given by (\ref{eq:rh2}), hence the radiation density $\rho
_{rad}=aT^{4}$ diverges as mass decreases, as in the usual picture
\cite{Pag76}. Thus mass loss increases indefinitely, and the central
mass vanishes as it's energy is carried
away by radiation; this process completes in a finite proper time (\ref%
{eq:lifetime}), as recorded by the central fluid.\newline

The central issue that determines the outcome is,
\begin{itemize}
  \item \textit{Where does the radiation go}? The answer is that most of it ends up in the spacelike future singularity rather than at infinity.
\end{itemize}
Because this radiation ends up at the central singularity, it does not carry energy or information to infinity; consequently the Hawking radiation process cannot make the central mass evaporate. \\

These issues will be considered after looking in more detail at the relevant processes (Section \ref{sec:revised_basic_process}). The outcome (Section \ref{sec:main}) depends crucially on the relevant domains (section \ref{sec:option2_domains}).

\subsection{The basic process}
\label{sec:revised_basic_process}

Assume there is an emission domain ED
that lies just outside the timelike EMOTS surface, and that local
energy conservation governs the emission process (Figure 5).\newline
\begin{figure}[tbp]
\includegraphics[width=7in]{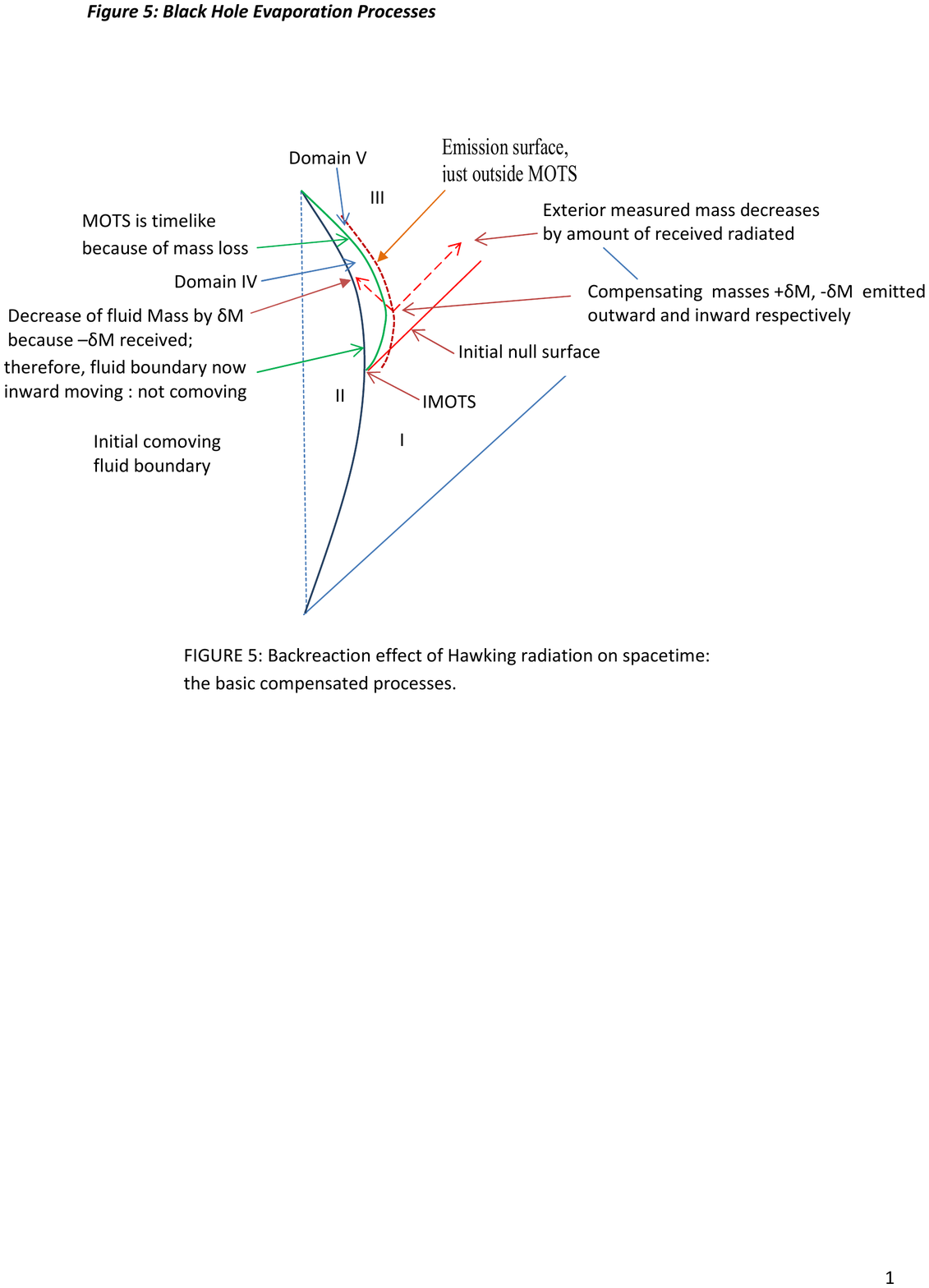}
\caption{\textbf{Figure 5}} \label{Fig 5}
\end{figure}

Imagine a time step $\delta t$ when radiation carries a chunk of \
(positive)\ energy $\delta E$ outwards and (because of energy conservation)\
a compensating chunk of negative energy $-\delta E$ inwards from the
emission domain. This can be understood as due to tunneling: as explained by Parikh and Wilczek,
\begin{quote}
``\emph{According to this picture, the radiation arises by a process similar
to electron-positron pair creation in a constant electric field. The idea is
that the energy of a particle changes sign as it crosses the horizon, so
that a pair created just inside or just outside the horizon can materialize
with zero total energy, after one member of the pair has tunneled to the
opposite side... energy conservation plays a fundamental role: one must make
a transition between states with the same total energy, and the mass of the
residual hole must go down as it radiates}'' \cite{ParWil00}.
\end{quote}

Seen from the exterior of the star, this causes the mass measured to
decrease, because positive mass $\delta m$ has been carried away as
radiation outwards and registered in the exterior:
\begin{equation}
m_{ext}(t+\delta t)=m_{ext}(t)-\delta m.  \label{eq:mloss1}
\end{equation}%
This also causes the EMOTS surface $r=2m$\ to move inward towards the centre
(as it contains less mass). The ingoing radiation carries negative energy -$%
\delta m$ to the interior, which then decreases the mass of the collapsing
fluid sphere in such a way that
\begin{equation}
m_{fluid}(t+\delta t)\ =m_{fluid}(t)-\delta m.  \label{eq:mloss2}
\end{equation}%
These two processes are repeated continuously and cause the mass of the
fluid sphere to die away towards zero, with the EMOTS therefore decreasing in
size towards zero. Following Feynman, the process can alternatively be
thought of as positive mass $\delta m$ traveling backwards in time from the
fluid surface to the EMOTS, and then being re-emitted there as positive mass $%
\delta m$ which is radiated outwards in the future direction of time towards
the exterior domain. Thus mass is transferred from the fluid to the exterior
by a quantum radiation process. This view is supported by Visser's detailed
calculations; he states

\begin{quote}
``\textit{What is the energy source for the Hawking radiation? The infalling
particles have negative energy as seen from outside the apparent horizon. In
the case of general relativity black holes the infalling particles serve to
reduce the total mass-energy of the central object, and it is ultimately the
total mass-energy of the black hole that provides the energy emitted in the
Hawking flux.}''  \cite{Vis01}.
\end{quote}
This is possible only because of the energy conservation emphasized by Parikh and Wilczek: there are ingoing and outgoing particles tunneling in opposite directions, and both channels -- particle and anti-particle tunneling -- contribute to the rate of the Hawking process \cite{ParWil00}.\\

For those who do not like the particle picture, one can calculate the vacuum expectation value of the energy momentum tensor of the relevant field. This has been done in the case of a two-dimensional model of the collapse of a shell of matter by Davies, Fulling and Unruh \cite{DavFulUnr76}. They find,
\begin{quote}
``\emph{In the collapse model, the results support that picture of black-hole evaporation in which pairs of particles are created outside the horizon (and not entirely in the collapsing matter), one of which carries negative energy into the future horizon of the black hole, while the other contributes to the thermal flux at infinity}''
\end{quote}
which fully agrees with the above picture. Furthermore they state
\begin{quote}
  ``\emph{The flux of energy is given by two components. Near infinity it is dominated by an outward null flux of energy. Near the horizon, however, it is a flux of negative energy going into the horizon of the black hole. This negative energy flux would presumably cause the area of the horizon to shrink at a rate consistent with the energy flux observed at infinity}.''
\end{quote}
Thus we can think of there being an emission domain just outside the horizon, which emits positive energy radiation going outwards and negative radiation going inwards.\\

Clifton \cite{Cli08} has calculated the stress tensor
$T_{ab}^{(out)}$ of the outgoing radiation for the case of a locally
defined horizon. He also finds ingoing as well as outgoing
radiation: because of the energy conservation condition, there is an
opposite inward energy flux with the stress tensor $T_{ab}^{(in)}$
of the same magnitude but with opposite sign as the outgoing
radiation. At the emission event,
\begin{equation}\label{eq:sum_stress}
  T_{ab}(out)|_H + T_{ab}(in)|_H=0.
\end{equation}
(see Section \ref{eq:findES} for further discussion).  It is the sum
stress tensor
\begin{equation}\label{eq:stress_sum}
T_{ab}(sum):=T_{ab}(out)+T_{ab}(in)
\end{equation}
that will be the source of the semiclassical backreaction causing focussing or defocusing of null geodesics, and depositing energy either in the fluid, at infinity, or at the future singularity. Like Davies et al, he emphasizes that the emission takes place a little outside the horizon. \\

Thus the main points are,

\begin{itemize}
\item Through the Hawking process, radiation carrying compensating masses $%
+\delta m$, $-\delta m$ are emitted outward and inward respectively in the
emission domain $ED$, which lies just outside the EMOTS, which is defined to
have zero external null convergence: $\theta _{+}=0$. This compensated
process is required in order that energy is conserved.

\item The exterior measured mass of the central fluid decreases by the
amount $+\delta m$ of received radiated in the exterior domain of the
emission surface, because this received energy indicates that a decrement
has taken place in the interior mass. However the issue is where this
radiation is deposited. To clarify this we need to carefully delineate the
causal structure of the situation (see the next section).

\item There is a decrease of fluid mass by $\delta m$ because of the amount
of energy $-\delta m$ received by the fluid from the emission surface.
Therefore, the fluid boundary is inward moving rather than comoving, as mass
is progressively chipped away from the outer edges of the fluid sphere.

\item Because this central mass has thus decreased through this mass loss,
the EMOTS, given by $r_{EMOTS} = 2m(v)$, is timelike: it continually moves inwards as the Schwarzschild
radius of the central mass decreases.
\end{itemize}
Thus the EMOTS surface is no longer the same as the event horizon: it lies in the interior.

\subsection{The emission region and the idealized emission surface}
\label{sec:revised_emission}
The key point is that the Hawking Radiation process is a local
process, and therefore that radiation is emitted near the
time-varying timelike EMOTS surface. We develop this in stages.

\subsubsection{Local radiation emission at a MOTS}
A detailed calculation showing radiation is emitted as a consequence of the local existence of a trapping surface  has been given by Visser \cite{Vis01},
based on use of Painlev\'{e}-Gullstrand coordinates
\begin{equation}
ds^{2}=-[c(r,t)^{2}-v(r,t)^{2}]dt^{2}-2v(r,t)dtdr+dr^{2}+\
r^{2}(d\theta ^{2}+\sin ^{2}\theta d\phi ^{2}).
\label{eq:Pain_Gull}
\end{equation}%
The apparent horizon is located at $r_{H}(t)$ given by
\begin{equation}
c(r_{H}(t),t)=|v(r_{H}(t),t)|.  \label{eq:Horiz}
\end{equation}%
On choosing
\begin{equation}
c(r,t)^{2}-v(r,t)^{2} = e^{-3\Phi(t,r)}[1 -
\frac{2m(t)}{r}],\,\,v(r,t) = e^{-3\Phi(t,r)}
[\sqrt{\frac{2m(t)}{r}}]\label{eq:mass_t}
\end{equation}%
(see \cite{NieVis06}), this shows that
\begin{equation}
r(t)|_H = 2m(t). \label{eq:horizon_mass}
\end{equation}%
The surface gravity is represented by a quantity $g_{H}(t)$ defined
by
\begin{equation}
g_{H}(t)=c_{H}\frac{d}{dr}[c(r,t)-|v(r,t)|]_{H}
\label{eq:surf_grav}
\end{equation}%
where the subscript $H$ means evaluated at the horizon
(\ref{eq:horizon_mass}). In the static case, the surface gravity is
\begin{equation}
\kappa =\frac{g_{H}}{c_{H}}.  \label{eq:kappa}
\end{equation}%

Using an Eikonal approximation for a quantum field $\varphi (r,t)$
written in terms of a rapidly varying phase times a slowly varying
envelope, Visser shows that outgoing modes occur with
\begin{equation}
\varphi (r,t)_{out}=\frac{N_{out}}{\sqrt{2\omega
}r_{H}}[r-r_{H}]^{\pm i\omega /\kappa }\exp \left\{ \mp i\omega
t\right\}   \label{eq:out_modes}
\end{equation}%
where $N_{out}$ is a normalisation factor. The calculation is based
solely in quantities evaluated at or near the apparent horizon,
which is located at the MOTS 3-surface.\newline

Consequently \cite{Vis01}, independently of anything related to
global event horizons,
\begin{quote}
\emph{\textbf{Hawking radiation emission}: Hawking radiation is
present under the circumstances of a slowly evolving timelike
apparent horizon, with amplitude varying as} $1/r_{H}$. \emph{It
will not  be emitted if the apparent horizon is spacelike}.
\end{quote}
It depends on the surface gravity $\kappa $ of the apparent horizon;
and in the context we are concerned with, that horizon is indeed slowly varying.%
\newline

An alternative detailed calculation showing that radiation emission
is a local process associated with a local horizon has been given by Parikh and Wilczek \cite%
{ParWil00}, based on the idea of Hawking radiation as tunneling;
this has developed further by Clifton \cite{Cli08}. The conclusion
is supported by other papers, including \ Nielsen (\cite{Nie08}:
Section 8) and Peltola \cite{Pel09}, who gives a local derivation of
Hawking radiation based in Bogoliubov transformations. Also
Paranjape and Padmanabhan  show that

\begin{quote}
``\emph{The presence of a flux of particles at large distances is
governed entirely by the local dynamics (in outgoing time u) of the
collapsing object. A flux will arise even in the situation where the
object starts at some radius $R_0$ and collapses to a final smaller
radius $R_f > 2M$ without forming a black hole, although the
spectrum will not be thermal}'' \cite{ParPqad09}.
\end{quote}
Parikh and Wilczek emphasize ``\emph{Note that only local physics
has gone into our derivations}'' \cite{ParWil00}. Clifton's paper
shows that the radiation emanates not from the MOTS\ 3-surface
itself, but from just outside that surface.\\

Taken together, the case is strong that Hawking  radiation is governed by the local existence of a trapping surface rather than by the existence of an event horizon, which by its nature is globally defined.  The domain where the radiation originates will be called the Emission Domain.

\subsubsection{Timelike or spacelike?}\label{sec:keypoint}
Now a key issue is how this radiation emissions depends on the causal nature of the MOTS 3-surface. Hawking radiation can
certainly happen for a null surface: that is the content of Hawking's original paper \cite{Haw73}). Can it happen if the apparent horizon is
timelike or spacelike?\\

The claim here is that this would radiation emission not occur if the MOTS surface were spacelike: it would not
then have required separation properties.
\begin{itemize}
  \item If we use the tunneling description \cite{ParWil00}, the MOTS must be timelike else there is no surface to tunnel through.
  The very concept of tunneling depends on the implicit assumption that the trapping surface causing the emission must be timelike
  so that two sides (``inside'' and ``outside'') can be defined.
  \item The concept of scattering involved in use of Bogoliubov transformations assumes scattering is off a timelike world tube.
\end{itemize}
Hence for the Hawking radiation emission to occur, the trapping surface must be a timelike MOTS.\\

Putting this together, we have a key conclusion:
\begin{quote}
\textbf{\emph{Keypoint: The Emission Domain}} \emph{Compensating ingoing and outgoing
radiation is emitted by an emission domain which lies just outside a timelike
MOTS 3-surface. }
\end{quote}
This will be the basis for the analysis that follows.\\

How do we determine its limits? The emission process can be represented in two ways (see the Appendix)
\begin{itemize}
  \item Virtual particles become real due to tunneling through the timelike horizon, depending on events in a volume $\delta V$ but conceptually occurring at a point $E$ in $\delta V$
  \item This result in changes in the energy-momentum tensor in a local volume: a zero tensor (on the input surface) to sum in and out (output surface)
\end{itemize}
Emission occurs in $\Delta V$ if $T_{ab}|_{(in)} = 0$ but $T_{ab}|_{(out)} \neq 0$.The Emission Domain is defined to be the smallest domain where this is true.

\subsubsection{Finding the Emission Surface}\label{eq:findES}
In order to consider how the location of emissions events relates to spacetime domains, it is useful to define an emission surface (ES) that represents the central location of the emission process. It will lie within the emission domain and is intended to represent the epicentre of emission. \\

How can one determine a unique locate for the ES? The energy-momentum-stress tensor gives the energy and 4-momentum
crossing a surface element $dS^{a}$ by the relation
\begin{equation}\label{eq:tabsb}
T^{a}=T_{ab}dS^{b}.
\end{equation}
The associated energy measured crossing the surface by an observer
moving with 4-velocity $u^{a}$ is
\begin{equation}\label{eq:deltaE}
\Delta E=u^{a}T_{a}=u^{a}T_{ab}dS^{b}.
\end{equation}
and the total radiation across a sphere $S(t,r)$ will be given by
\begin{equation}\label{eq:radnsphere}
E=\int_{S(t,r)}u^{a}T_{ab}dS^{b}.
\end{equation}
In the context of the Hawking evaporation, we have inwards and
outwards components
\begin{equation}\label{eq:hawplusminus}
T_{ab}=T_{ab}^{(in)}+T_{ab}^{(out)}
\end{equation}
(cf. \cite{ParWil00},\cite{Cli08}) so
\begin{equation}\label{eq:radnsum}
T_{a}=\left( T_{ab}^{(in)}(r,t)+T_{ab}^{(out)}(r,t)\right) dS^{b}.
\end{equation}
The associated energy is
\begin{eqnarray}\label{eq:radnsumplus}
E|_{S(t,r)} &=&E^{(in)}|_{S(t,r)}+E^{(out)}|_{S(t,r)}
\end{eqnarray}
where
\begin{eqnarray}\label{eq:ein}
E^{(in)}|_{S(t,r)} &:=&\int_{S(t,r)}u^{a}T_{ab}^{(in)}(r,t)dS^{b},\,\,\\
E^{(out)}|_{S(t,r)} &:=&\int_{S(t,r)}u^{a}T_{ab}^{(out)}(r,t)dS^{b}.
\end{eqnarray}

Given these definitions, the idea now is that if the outward flux
dominates over the inwards flux, you are outside the emission
surface; if the inward flux dominates over the outwards flux, you
are inside the emission surface; and if they are equal, you are at
the emission surface.  Thus the emission surface $S(t,r)_{ES}$\ is
determined by seeing when equality happens.
\[
\begin{tabular}{|l|l|}\hline\hline
Inside the Emission surface &
$|E^{(in)}(S(t,r))|\,>\,|E^{(out)}S(t,r)|$ \\\hline On the Emission
surface & $|E^{(in)}(S(t,r))|\,=\,|E^{(out)}(S(t,r))|$
\\\hline
Outside the Emission surface & $|E^{(in)}(S(t,r))|\,<\,|E^{(out)}(S(t,r))|$\\\hline\hline%
\end{tabular}%
\]
\textbf{Table 2}: \emph{Determining the location of the Emission Surface}.\\

Hence: At given $t,$ vary the position of the surface inwards  and outwards to find the radius $%
r_{ES}(t)$\ such that
\begin{equation}\label{eq:surfaceofE}
|E^{(in)}(S(t,r_{ES}(t)))|=|E^{(out)}(S(tr_{ES}(t)))|
\end{equation}
It does not matter what surfaces $\left\{ t=const\right\} $ are used
to do this because this condition depends only conditions on the
2-surfaces $S(t,r).$\\

Which 4-velocity should be used in this process? -- the 4-velocity
of the world tube $S(t,r_{ES}(t))$ so determined! That is, the
vector field $\partial/\partial t$ tangent to the world lines
$x^a(t) = (t,r_{ES}(t),\theta_0,\phi_0)$) for all
$\theta_0,\phi_0$). This will have to be calculated in an iterative
manner to obtain a consistent solution.\\

In what follows, we will use the emission surface ES as a proxy for the emission domain ED. This will generally enable us to understand what is going on well and enables us to consider relevant domains in a precise way. We will need to make the distinction between the ES and ED only in considering the possibility that radiation escapes to infinity, see Section \ref{sec:mass loss}. Until the referring to the ES will suffice for our purposes.

\subsection{The Nature of the Spacetime}\label{sec:main}
We now have all the pieces needed to determine where Hawking radiation emission takes place in a astrophysical blackhole setting.

\subsubsection{The Dynamical Horizons}\label{sec:dynamical1}
We saw in Section \ref{sec:dynamical} (Table 1) that the sign of $\partial \theta _{(\ell )}/\partial n^{a}$ determines whether the MOTS 3-surface is timelike or spacelike. Adding the conclusion stated in the Keypoint (Section \ref{sec:keypoint}) that the MOTS must be timelike for Hawking radiation emission to occur, any MOTS that emits radiation must be an EMOTS and the situation is as summarised in the following table:\\

$%
\begin{array}{|l|l|l|l|l|l|}
\hline \hline \text{Horizon} & \theta _{(\ell )}
&
\partial \theta _{(\ell )}/\partial n^{a} & \text{Nature:} &
\text{Radiation?} \\ \hline \hline \text{EMOTS}  &
\theta _{(\ell )}=0 & \partial \theta
_{(\ell )}/\partial n^{a}>0 & \text{timelike} & \text{emits radiation} \\
\hline \text{OMOTS}  & \theta _{(\ell )}=0 &
\partial \theta _{(\ell )}/\partial n^{a}<0 & \text{spacelike} &
\text{emits no radiation}
\\ \hline \hline
\end{array}%
$\newline \\

\textbf{Table 3}: \emph{Outgoing null geodesic divergences according to domain}.\\

This enables us to determine where radiation emission occurs. Recall
the causal domains (Figure 5). There is an outer spacelike OMOTS
3-surface and an inner timelike EMOTS 3-surface, each of which is a
dynamic horizon comprised of 2-spheres such that
$\hat{\theta}_{+}=0$. These two 3-surfaces intersect in the inmost
marginal outer trapped 2-surface (IMOTS) at the fluid surface, which
is where the black hole properties originate.

\begin{itemize}
\item The outer OMOTS 3-surface is spacelike and goes to infinity, indeed it
defines infinity; the event horizon lies inside it, because the
event horizon is null. It is not associated with radiation emission
because it is spacelike; however it is associated with the
externally measured mass and angular momentum (see \cite{AshKri02},
\cite{Dreetal03}, and references therein).

\item The inner EMOTS 3-surface is timelike, and lies just outside the collapsing fluid; it is this surface
associated with the emission of radiation. More specifically: rather
than being associated with the event horizon $r=2m_{0},$ which does
not know about the mass loss, the radiation emission surrounding a
collapsing star of initial mass $m_{0}$ occurs at an emission
surface ES\ just outside the EMOTS\ surface (\cite{ParWil00},
\cite{Vis01},  \cite{Cli08}) which lies just outside the fluid
surface. This surface knows about the mass loss and accordingly
changes in time (it will be located at $r_H=2m(t)$, see
(\ref{eq:horizon_mass})). It is this surface that is represented in
Figure 5.
\end{itemize}
This is rather surprising because the emission in the primordial
black hole case --  that is, the extended Schwarzschild solution with no matter in it --- is understood to be from the vicinity of the event horizon at $r=2m_0$, and it seems natural to assume this will be the case in the collapsing star scenario. If we allow for the radiation taking place on a MOTS rather than the event horizon, as motivated in the last section, it therefore seems obvious it would be emitted by the OMOTS surface that lies close to the event horizon. But this is not the case: the OMOTS surface is spacelike and so emits no radiation. It is emitted from the timelike EMOTS surface, close to the fluid. Why is this so? How can the source be so displaced from what we expect?\\

\begin{figure}[tbp]
\includegraphics[width=7in]{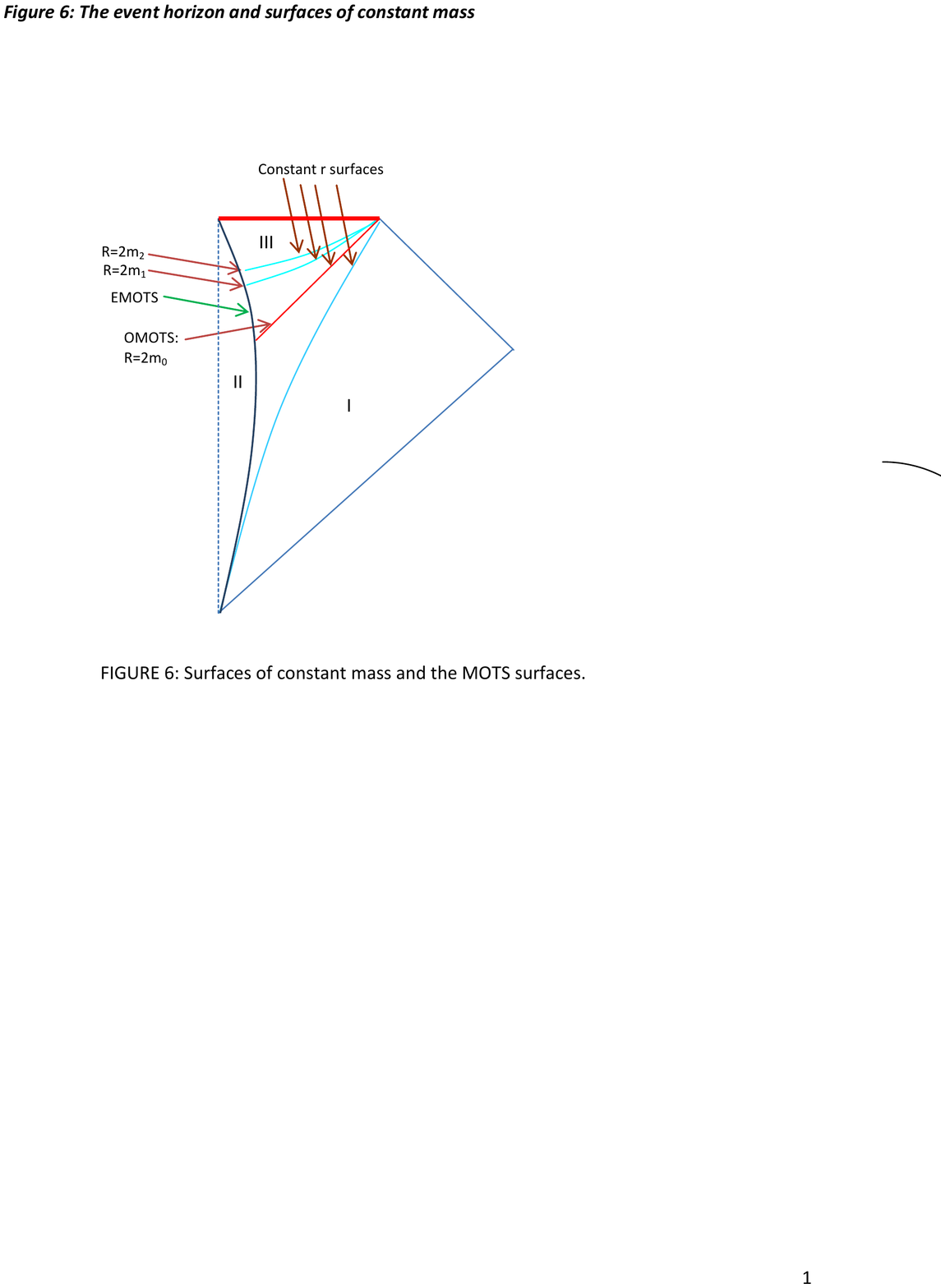}
\caption{Figure 6} \label{Fig6}
\end{figure}
To explain this further, consider Figure 6, a conformal diagram of collapse of fluid sphere of initial mass $m_0$.  For the moment we neglect the CMB effects shown in Figure 2, so this is the same as Figure 1, but now some surfaces $\{r=constant\}$. Because this is a part of the maximal extension of the vacuum Schwarzschild solution, and $r<2m_0$ in this domain, these surfaces are spacelike are shown in Domain III \cite{HawEll73}. The OMOTS surface $r = 2m_0$ is of course also a surface of constant $r$. These surfaces are timelike Domain II; it is the discontinuity in their nature that lead to their divergence in the diagram, in effect generated by the bifurcate Killing horizon that exists in the maximal solution \cite{Boy69}, even though it is not actually present in this spacetime.\\

Now the key feature is that $r=2m$ is an entire 3-surface; so the question is, from where on that surface does the radiation arise? Does it come from the entire surface, hence including locations indefinitely far from the star? Or does it come from domains close to the collapsing star? \\

The solution has already been given above: it will be emitted near the fluid, that is at the EMOTS surface (see Figure 1) where the sign of the outgoing null geodesic expansion changes. Hence the emission location in the 3-surface $r=2m$ is close to the $I_{MOTS}$ 2-sphere where $r=2m$ intersects the EMOTS at the surface of the fluid surface. In fact the emission takes place \emph{because} this 2-sphere is in the EMOTS, rather than because it is near the OMOTS surface $r=2m_0$. The point is that the OMOTS surface and the EMOTS surface come into existence simultaneously when the negative $\hat{\theta}_+$ Domain III comes into being at the $I_{MOTS}$ 2-sphere. But the one that causes the radiation is the EMOTS, located at the fluid surface. The radiation from this surface starts at the time $t_H$ when the horizon forms, which is why the radiation is associated with the entire null OMOTS.\\

We can see the rationale by considering where the surfaces $r=2m(t)$ lie. Specifically how do the surfaces $r=2m(t)$ intersect the surface of the fluid  as $m(t)$ decreases with time? Remember that the surfaces $r=constant$ are spacelike for $r<2m$ (Figure 6). As radiation is emitted and $m$ decreases from its initial value $m_0$, it will take on a new value $m_1<m_0$ which will have associated with it a new trapping radius $r = 2m_1$. Viewed in the Diagram 6 for a Schwarzschild mass $m_0$, where on the surface $r = 2m_1$ will emission take place? It has to be at a MOTS; and the only place the spacelike surface $r = 2m_1$ will intersect a MOTS is at the fluid outer surface. It intersects this surface in a 2-sphere $S_{MOTS}(t_1)$ at a later time than the OMOTS surface $r=2m_0$ does. So if we take the $r$ value $r = 2m(t)$ as determining where radiation is emitted then, this has to happen where the surface intersects a marginally trapped surface. But it does not intersect the OMOTS surface; rather it intersects the EMOTS surface (that is,the fluid boundary) as shown in the diagram, and that will have to be the source of the radiation at the later time. \\

This is the clear implication of the \emph{Keypoint} (Section \ref{sec:keypoint}) and the
resulting Table 3. This conclusion will remain true when we take into account the
effect of the CMB on the causal domains (see Figure 2).

\subsubsection{Revised domains}\label{sec:domains_revise1} \
Consequent on the effects of the Hawking radiation, there are two
new domains that need to be recognized in the causal structure (see
Figure 7; they were also indicated in Figure 5).
\begin{itemize}
\item \textbf{Domain V}: the emission surface is outside the EMOTS surface. The region between them is Domain V. It is full of negative energy radiation streaming towards the EMOTS.
\item \textbf{Domain IV}: this negative energy density defocuses radiation coming from the EMOTS surface, which (see Figure 3) in the classical case is the fluid surface. Thus there is now a region just outside the fluid surface which has positive outgoing null expansion.  Hence the Hawking radiation moves the EMOTS surface off the fluid boundary and into the surrounding vacuum. The region thus created is Domain IV. It is full of negative energy density radiation streaming inwards
\end{itemize}
These domains are discussed more fully in the Section \ref{sec:option2_domains2}.

\subsection{The Main Hypothesis}\label{sec:main1}
We will now develop the consequences that follow from all this.

\subsubsection{The assumption and the outcome}
We make the key assumption that Hawking radiation is emitted
from just outside the EMOTS\ surface in Figure 2, which seems to be the
realistic view.
\begin{quote}
   \textbf{ Main Hypothesis}: The source of outgoing Hawking radiation is neither near
   the event horizon,  nor the outer trapping (OMOTS) surface: its
   location is near the timelike EMOTS surface close to the fluid
   surface.
\end{quote}
\begin{figure}[tbp]
\includegraphics[width=7in]{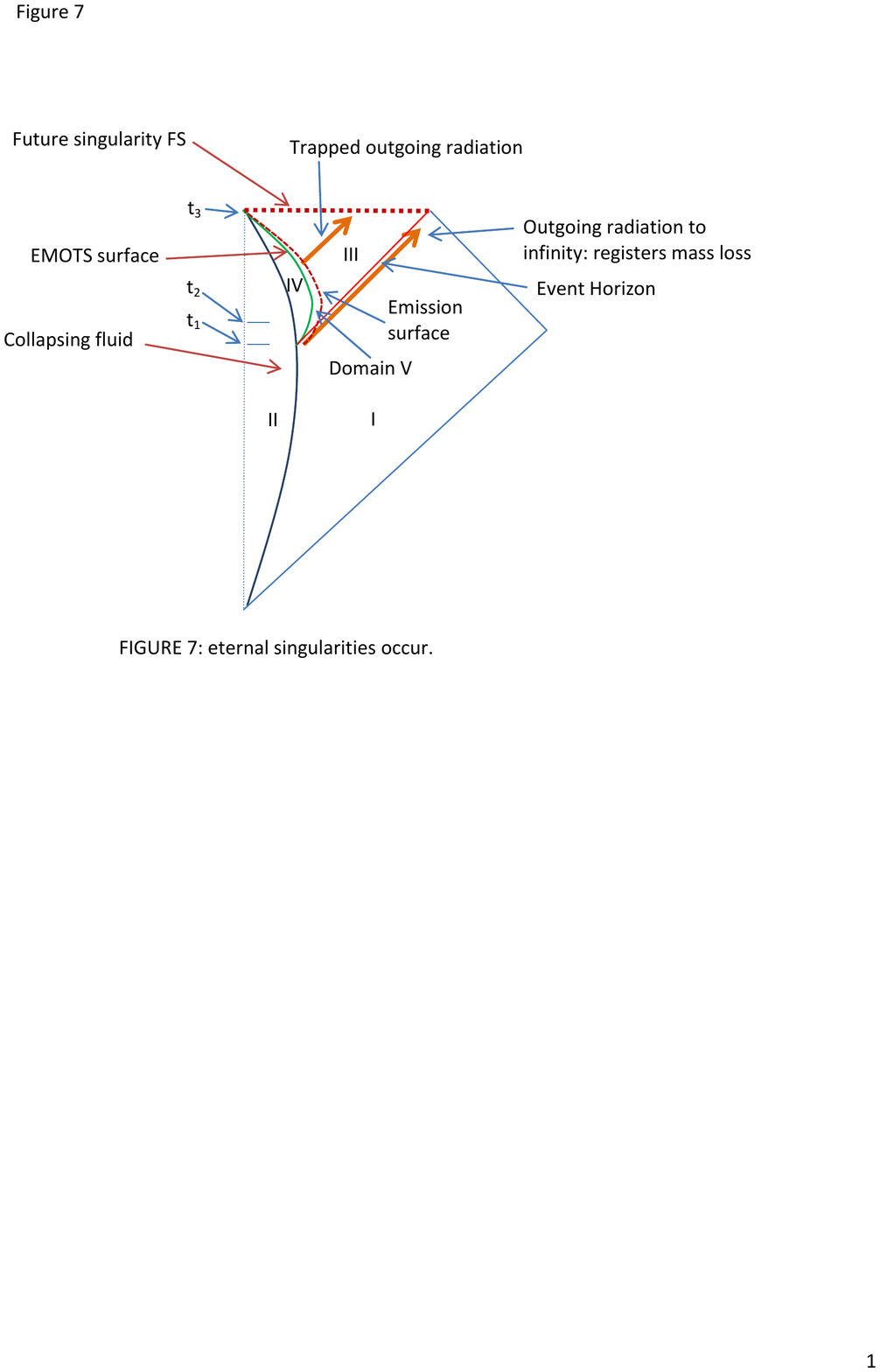}
\caption{Figure 7} \label{Fig7}
\end{figure}
The broad nature of the resulting spacetime is shown in Figure 7; the relevant domains are show in Figure 8, indicating the sign of the outgoing null geodesic expansion $\hat{\theta}_+$. The changes in sign of $\hat{\theta}_+$ determine the location of the MOTS surfaces, and hence characterises where radiation will be emitted. \\

An immediate consequence is
\begin{quote}
   \textbf{Corollary}: Most of the Hawking radiation emitted during the collapse process does not end up at infinity: it ends up on the future singularity. Consequently the singularities  do not evaporate away because of outgoing Hawking radiation; that radiation ends up on the singularity, and so does not carry any mass or energy away to the exterior
\end{quote}
Why is this so? Because the EMOTS surface lies inside the event horizon (see
Figure 7), hence the Hawking radiation is trapped. In fact the outgoing Hawking radiation adds extra mass to the outer part of the singularity, balanced by the decrease in mass caused by the ingoing negative energy radiation. Overall the mass that will be measured externally is conserved.

\subsubsection{Fallback position} Suppose that the  Main Hypothesis was not true. Then the fallback position must be that the source of the radiation is the spacelike OMOTS surface. Then negative density ingoing Hawking radiation will occur in domain III, and cause defocusing there, opening up the possibility that the future of that domain is singularity free; there would be a competition between this radiation and the CMB to determine what happens. But positive density outgoing Hawking radiation would still occur in Domain VI, which would be a trapping region and would lie behind the event horizon.\\

 Thus the trapping of the outgoing radiation would still be true at least in Domain VI even if the radiation was emitted from the OMOTS rather than the EMOTS, as Figure 8 (which develops from Figure 2) makes clear. That radiation too would fall into the singularity because the OMOTS is spacelike. The details would be different, but the outcome is likely to be the same.

\section{Eternally Surviving Black Holes}
\label{sec:eternal}

To bring this all together: the key point in the whole argument is the locality of the Hawking radiation effect. Most of it is emitted at the EMOTS close to the central matter; this is far inside the event horizon, so that radiation is trapped. Thus astrophysical black holes radiate but do not evaporate; rather there is always a
remnant mass left behind. They do not pop out of existence due to evaporation by Hawking radiation\\

The broad outline of what happens is thus clear. It is still worth examining the causal domains in more detail (Section  \ref{sec:option2_domains}), because that is required in order to consider the remaining big question: does any Hawking radiation escape to infinity, and if so, how much? That is considered in Section \ref{sec:mass loss}.

\subsection{The Trapped Surfaces and the Domains}
\label{sec:option2_domains}
The dynamic processes taking place revolve around the concept of a
dynamical horizon (Section \ref{sec:dynamical}).

\subsubsection{Trapping Surfaces and the Event Horizon}\label{sec:option_horizons}
There is an entire literature on \textquotedblleft isolated
horizons'\ or dynamic horizons`' (\cite{AshKri02}, \cite{Dreetal03})
that refers to spacelike horizons like the OMOTS. The work above on
local radiation emission (Section \ref{sec:revised_emission}) refers
to the timelike EMOTS. Both occur in these solutions, but in
different domains. With that understanding it all fits together: The resulting broad causal structure is presented in Figure 8, which is very similar to the classical astrophysical collapse situation (Figure 2), but with extra structure added. \\

The key features shaping the causal structure are the two MOTS surfaces, whose location is determined by backreaction effects (Figures 2 and 8).
\begin{quote}
\textbf{The Trapping surfaces}: \emph{
In astrophysical black holes there is an inner timelike EMOTS
3-surface and an outer spacelike OMOTS 3-surface that both originate
at the IMOTS 2-surface. Outgoing Hawking radiation is emitted at the Emission
Surface ES just outside the EMOTS, and mainly ends up at the future singularity.}
\end{quote}
The OMOTS starts at the IMOTS 2-sphere and then moves outwards, and
is therefore spacelike. It determines the scope of the singularity,
by bounding the trapped 2-surfaces in Domain VI; hence it determines
the location of the event horizon, which is generated by the last
outgoing null geodesics that just fail to hit the singularity. It
lies outside the initial outgoing null surface that passes through
the IMOTS\ 2-surface.

\subsection{Causal structure}\label{sec:option2_domains2}
\label{sec:revised_domains}
To see how this works out we must carefully delineate the relevant domains and their boundaries.

\subsubsection{Domains}
To understand the resulting causal structure, we need six domains
shown in Figure 8.
\begin{figure}[tbp]
\includegraphics[width=7in]{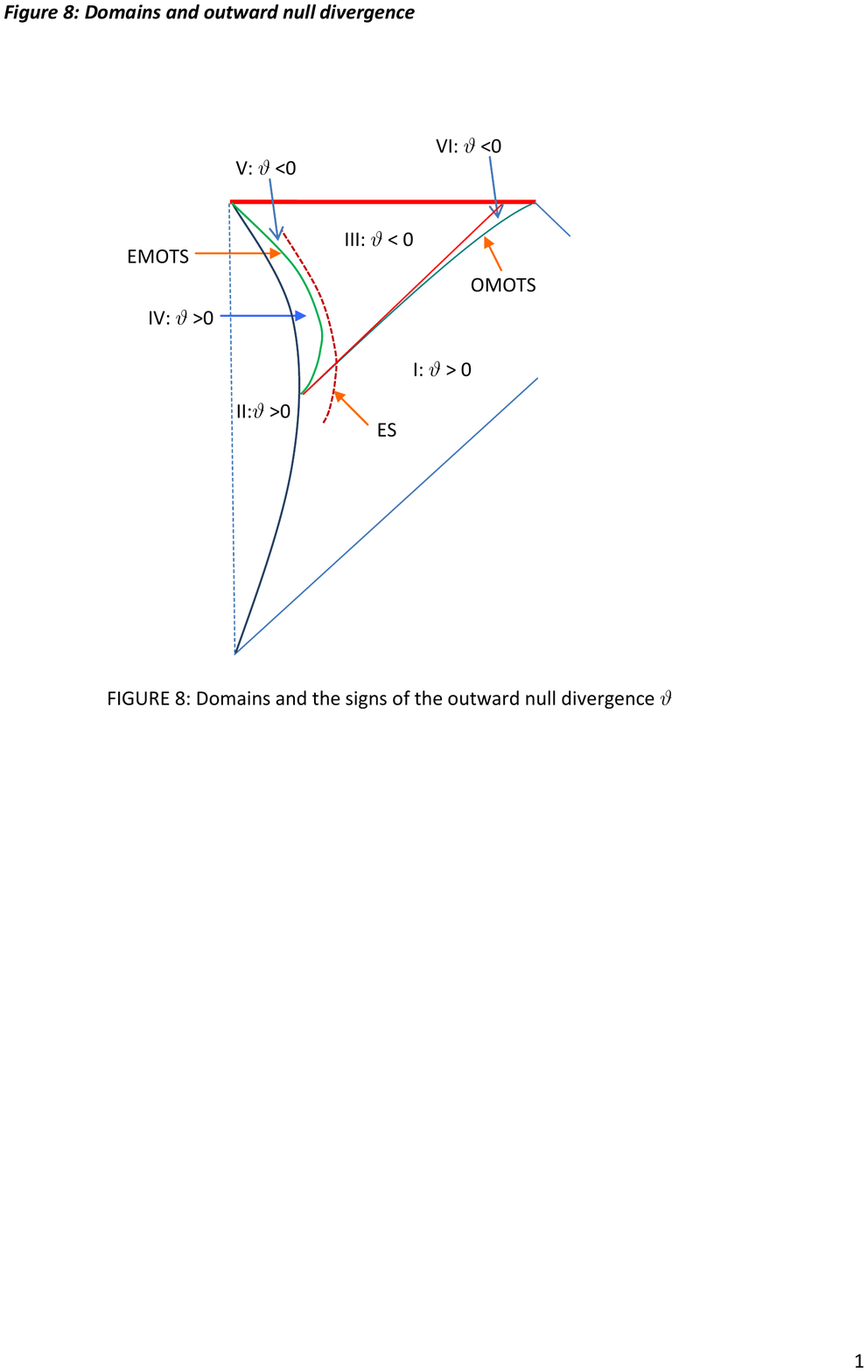}
\caption{Figure 8} \label{Fig8}
\end{figure}
Compared with Figure 2, two extra domains are added (Section \ref{sec:domains_revise1}).
Thus the domains now are
\begin{enumerate}
\item \textbf{Domain I}, an exterior vacuum solution for a mass $m_{0}$, bounded on the outer sides by past and future null infinity.
It is bounded on the inner side by the fluid initially, and then by the Initial
Null Surface  at $r=2m_{0}$ (this may or may not be an event horizon;
that is to be determined later). This is an empty region, so it is a
Schwarzschild solution and we can use Schwarzschild results in this domain,
which is why we can initially identify the Initial null surface as being at $r=2m_{0}$
(this will be modified later due to backreaction effects).
\item \textbf{Domain II}, infalling matter of initial mass $m_{0}$, which we
take to obey the usual energy conditions. This can be a flat FRW dust model $(0<r<R_{S}(u))$ collapsing within Domain I, or a more complex fluid model.
We take this to include all infalling matter; that is, no further
shell of matter is falling in from outside this domain.
\item \textbf{Domain III}, the domain outside the Emission Surface (ES) and with
initial boundary the Initial Null Surface, filled with outgoing Hawking
radiation. This domain may or may not be a Vaidya outgoing radiation
solution. The energy conditions are satisfied here because the outgoing
radiation has a positive energy density.
\item \textbf{Domain IV}, the domain inside the timelike EMOTS 3-surface and outside
the collapsing fluid sphere, filled with ingoing Hawking radiation. This can
possibly be a Vaidya ingoing radiation solution. The energy conditions are
not satisfied here because the ingoing radiation has a negative energy
density.
\item \textbf{Domain V}, the small domain between the timelike EMOTS surface and the ES, with
ingoing Hawking radiation. The energy conditions are not satisfied here
because the ingoing radiation has a negative energy density.
\item \textbf{Domain VI}, the small region between the initial null surface
and the spacelike OMOTS surface which lies outside the Initial Null Surface.
\end{enumerate}

Where does the action start?\ At the innermost MOTS 2-surface where marginal
closed trapped surfaces first form:

\begin{itemize}
\item \textbf{IMOTS}: The key 2-sphere in the diagram is the innermost MOTS
2-surface. It is the initial MOTS\ 2-sphere that forms as the fluid collapses: there
are no MOTS 2-surfaces at earlier times.
\end{itemize}

\subsubsection{Separation surfaces}
The separation surfaces between these domains are as follows:

\begin{itemize}
\item \textbf{B12:} Domain I and Domain II are separated by the fluid boundary (the matter
surface MS), up to the time when the IMOTS forms. The mass $m_{0}$ within the surface is constant.

\item \textbf{B16:} Domain I\ and VI\ are separated by the spacelike OMOTS.

\item \textbf{B36:} Domain VI\ and III are separated by the Initial Null Surface, which
lies inside the ever increasing OMOTS.

\item \textbf{B24:} Domain II and Domain IV are separated by the matter surface,\ given by
coordinates $r=R_{S}^{+}(\tau)$ in Domain II. The mass $m_{0}(\tau)$\ within the surface is
decreasing with proper time $\tau $ along this surface.

\item \textbf{B45:} Domain IV and V\ are separated by the timelike EMOTS 3-surface,\ given by
coordinates $r=r_{MTS}^{+}(\tau)$ in Domain I\ and $r=r_{MTS}^{-}(\tau)$ in Domain II.
Every 2-sphere in this surface is a MOTS with $\hat{\theta}_{+}=0.$

\item \textbf{B35:} Domain V and III\ are separated by the ES, which lies outside the
ever decreasing EMOTS.

\end{itemize}

\subsubsection{The outgoing divergence values}\label{sec:divergence1}
\noindent The sign of the outgoing null vector expansion $\theta
_{(\ell )}$ (cf. Section \ref{sec:divergence} and Table 1; Section
\ref{sec:dynamical1} and Table 3) is  as shown in Table 4.

\begin{center}
$
\begin{array}{|c|c|c|} \hline \hline \text{Domain} & \theta
_{(\ell)} & \text{Matter content} \\\hline \hline \text{I} & \theta
_{(\ell)} >0 & \text{empty} \\ \hline \text{II} & \theta _{(\ell)}
>0 & \text{matter} \\ \hline \text{III} & \theta _{(\ell)} <0 &
\text{radiation} \\ \hline \text{IV} & \theta _{(\ell)} >0 &
\text{negative density radiation} \\ \hline \text{V} & \theta
_{(\ell)} < 0 & \text{negative density radiation}
\\ \hline \text{VI} & \theta _{(\ell)} <0 & \text{radiation} \\
\hline \hline
\end{array}
$%
\end{center}

\textbf{Table 4:} \emph{Domains, matter content, and signs of outwards normals}.\\

The consequent trapping regions are as indicated in Table 4.\\

$%
\begin{array}{|l|l|l|l|l|l|}
\hline \hline \text{Horizon} & \text{Separates:} & \theta _{(\ell )}
&
\partial \theta _{(\ell )}/\partial n^{a} & \text{Nature:} &
\text{Radiation?} \\ \hline \hline \text{EMOTS} & \text{IV and V} &
\theta _{(\ell )}=0 & \partial \theta
_{(\ell )}/\partial n^{a}>0 & \text{timelike} & \text{emits radiation} \\
\hline \text{OMOTS} & \text{I\ and VI} & \theta _{(\ell )}=0 &
\partial \theta _{(\ell )}/\partial n^{a}<0 & \text{spacelike} &
\text{emits no radiation}
\\ \hline \hline
\end{array}%
$\newline \\

\textbf{Table 5}: \emph{Outgoing null geodesic divergences according to domain}.\\

\noindent It is these tables that justify the correctness of the geometry shown in Figure  8, because they locate the MOTS surfaces.

\subsection{Final issues} \label{sec:revised_future_sing}

The basic conclusion is that much of the Hawing radiation does not
escape to infinity, much of the central mass is not radiated away as
in the usual picture. Thus the black hole mass does not decay to
zero:\  it decreases  to a finite positive value, which is the
asymptotic value $m_{final}$ of the mass of the black hole as
measured from outside.\newline

However we need to consider two final issues before going on to look
at implications. First, is defocusing of null geodesics in Domain V
a serious problem? Second, does any Hawking radiation escape to
infinity?

\subsection{Focussing of geodesics}
\label{sec:revised_focussing}
Now a key issue in these models that needs checking is whether or
not the outgoing null geodesics from the EMOTS 2-spheres get
focussed at a conjugate point at a finite affine parameter value, or
not. The issue is that while these originate with zero divergence at
the EMOTS, they then get defocussed in Domain V before they enter
Domain III; so when they enter Domain III, they are diverging. Will
they then get refocussed and develop conjugate points, that then
imply singularities will develop in the future? The answer is yes,
because of the CMB radiation that played a key role in Section
\ref{sec:cbr1}.\\

The outgoing geodesics from any MOTS\ 2-surface start with vanishing divergence:\ $%
\hat{\theta}_{+}=0$ (because that is what defines the MOTS). Now the outgoing null radiation with a stress tensor of the form
\begin{equation}\label{eq:null_stress}
T_{ab} = \rho k_a k_b, \,\, k^ak_a =0
\end{equation}
has no focussing effect on the outgoing geodesics because the term they contribute to (\ref{eq:null_ray}) will vanish. However in fact that stress tensor is not a solution of Maxwell's equations, and in any case a more
complex matter tensor will occur in practice for the outgoing Hawking
radiation \cite{Cli08}; this may tend to cause focussing.\\

But additionally, the CMB emitted from the hot big
bang era in the early universe (section \ref{sec:cbr1}) will enter domain III\ from domain I\ and provide a
focussing term in the Null Raychaudhuri equation
(\ref{eq:null_ray}):

\begin{quote}
\textbf{\emph{The converging effect of the CMB radiation}} \emph{The outgoing null geodesics from the MOTS
2-spheres experience focusing by ingoing CMB\ radiation .}
\end{quote}

Will the initial defocusing tendency of the outgoing null geodesics in Domain V
be compensated by the focussing effects of the CMB in Domain III, so that they develop conjugate points? The point to note is that if there were no refocussing the
geodesics would be unbounded; but then the CMB's focussing effect would have
unlimited affine parameter values in which to act in Domain III, and so
would probably eventually overpower any initial divergence that might occur
in Domain V.\newline

For a very rough estimate, suppose $\hat{\theta}(\lambda _{1})>0$ for some
value $\lambda _{1}$ of $\lambda .$ The null Raychaudhuri equation (\ref%
{eq:null_ray}) shows that
\begin{equation}
\frac{d\hat{\theta}}{d\lambda }<-R_{ab}k^{a}k^{b}.  \label{eq:null_ray_focus}
\end{equation}%
Hence for any domain $(\lambda _{1}<\lambda )$ where $R_{ab}k^{a}k^{b}>c>\
0, $ we will have $\hat{\theta}(\lambda )<c(\lambda -\lambda _{1})+\hat{%
\theta}(\lambda _{1}).$ Thus
\begin{equation}
\hat{\theta}(\lambda )<0\;\mathrm{for}\;\lambda >\lambda _{1}+\hat{\theta}%
(\lambda _{1})/c.
\end{equation}%
Hence a finite constant CMB term in Domain III\ will cause
refocussing, no matter how large the defocusing may be in Domain 5,
because  if we assume no refocussing occurred, then the focussing
effect (\ref{eq:null_ray_focus}) would act for arbitrarily large
affine parameter values in Domain III; and that would indeed cause
focusing, in contradiction with the assumption.\newline

Now the CMB will actually be decreasing in the far future, but on the other
hand its initial value will be far greater than $T_{CMB}|_{0}$. The first
point here is that for gravitational collapse of black holes that exist
today, it will have been larger than $T_{CMB}|_{0}$ at the time when they
started forming in the earlier universe. But additionally, and more
important, (\ref{eq:blueshift}) holds for the incoming CBR, which starts off
at an effective temperature $T_{CMB}|(\lambda _{1})>T_{CMB}|_{0}$\ at the effective finite infinity as it
approaches the black hole. Therefore it gets blueshifted without limit as it
approaches the Initial Null Surface in the Domain I\ Schwarzschild solution:
equation (\ref{eq:CMB_Temp}) holds with $(1+z)\rightarrow 0$ as $%
z\rightarrow -1$ (blue shifting!). Then
\begin{equation}
\frac{\nu (r)}{\nu _{\infty }}=\sqrt{\frac{1}{1-2m/r_{R}}}\rightarrow \infty
\;\mathrm{as\;}r\rightarrow 2m,
\end{equation}
so the CMB\ radiation density that will be the source term on the right of (%
\ref{eq:null_ray_focus}) is apparently growing without limit as the
radiation approaches the Initial Null Surface. This high density of
radiation will enter domain III. Thus even if $\hat{\theta}_{1}>0$ at some
parameter value $v_{1}$ on the outgoing geodesic, there surely will be a
finite time $t_{2}>t_{1}$ such that $\hat{\theta}_{2}>0$ and conjugate
points will occur.\newline

But this applies to every MOTS 2-sphere $S(v)$ in the EMOTS 3-surface. Thus
it will apply at some time $v_{1}$ along the EMOTS\ 3-surface; every later
MOTS 2-surface for times $v_{2}>v_{1}$ will lie in the trapped future of $%
S(v_{1})$ and hence the radiation emitted at $v_{2}$ will necessarily end up
at the future singularity that occurs in the future $J^+S(v_{1})$ of $S(v_{1})$, \ and so
does not reach infinity. Then the boundary of the future of the MOTS is compact, and
by Penrose' argument (\cite{Pen65}, \cite{HawEll73}) singularities will occur in the
future. Furthermore, the outgoing Hawking radiation will to a good approximation have an
energy momentum tensor (\ref{eq:null_stress}) with $\rho_{CMB} > 0$; when conjugate points
occur, because of energy conservation along the null geodesics this density will diverge and
a Ricci singularity will occur there. Penrose's argument shows that a singularity occurs
somewhere, but not necessarily that it will occur in the causal future $J^+(S_{MOTS})$ of
the MOTS 2-sphere; this argument show that this will indeed be the case.\\

Thus a consequence of the existence of the cosmic CMB is that the
Hawking radiation is trapped and causes a future singularity to
occur. There is a nice symmetry about this result:\ on the classical
picture, the existence of the CMB radiation confirms that past
cosmological singularities are unavoidable (\cite{HawEll68},
\cite{HawEll73}); we now see it also confirms the inevitable
existence of future singularities in the collapsing black hole
context.

\subsection{Radiation escape and mass loss}\label{sec:mass loss}
The final question is whether any Hawking radiation escapes and gets to infinity? An idealised model of the situation is given in
Figure 9.\\

\begin{figure}[tbp]
\includegraphics[width=7in]{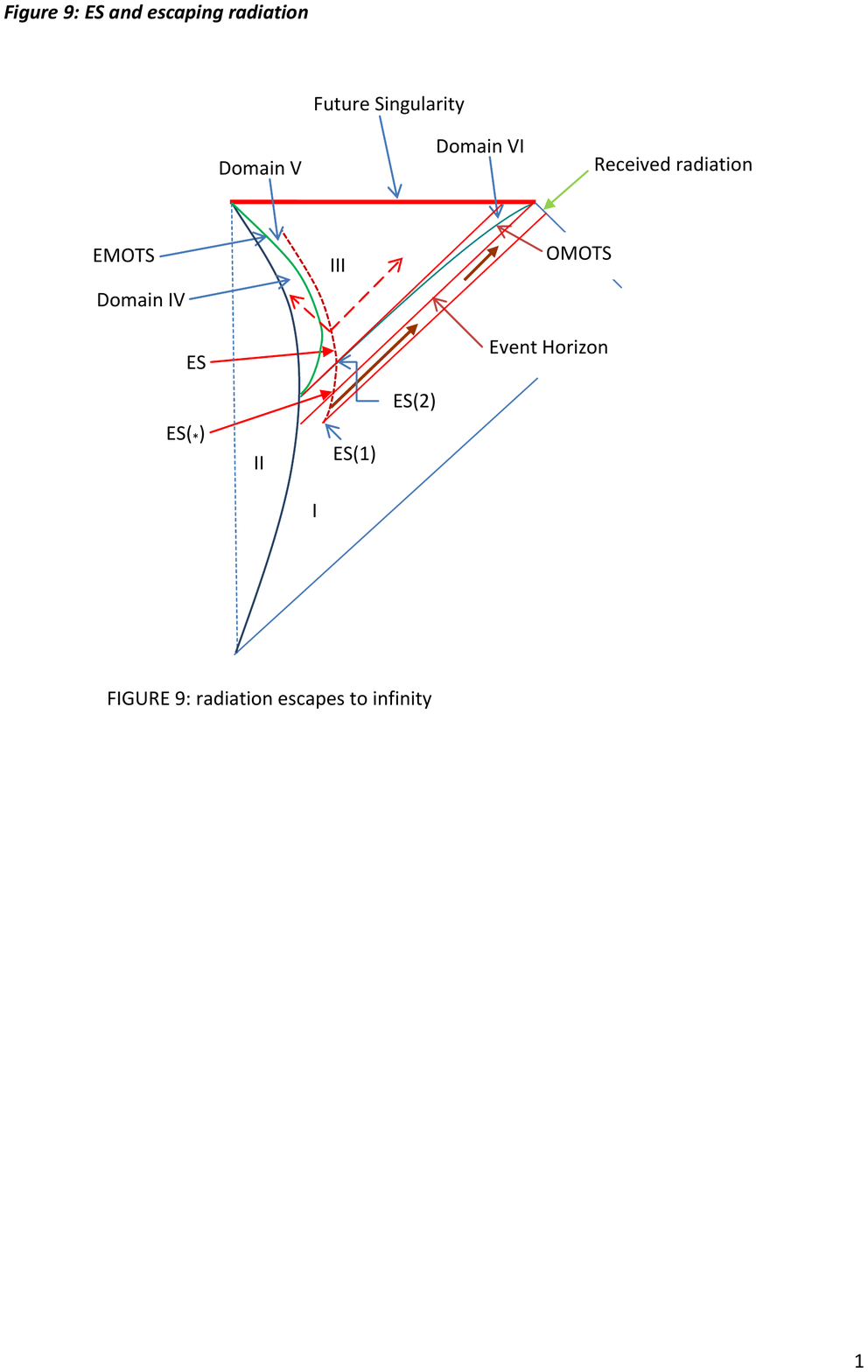}
\caption{Figure 9} \label{Fig9}
\end{figure}

The Hawking radiation is generated by the emission domain ED surrounding the emission surface ES, which lies
close to the EMOTS, but crucially, it lies a little outside the
EMOTS, as discussed above. If this was not the case, no Hawking radiation
would escape to infinity:\ it would all be trapped within the
initial null surface. Most outgoing radiation from the ED  does not
reach infinity; it ends up on the future spatial singularity FS, as is clear from Figure 8.
Therefore no corresponding change of mass is recorded at infinity by
this radiation. Even if some of this radiation went through the
center and out the other side \cite{BirDav84},
it will still end up on the future singularity rather than infinity. \\

Consider the idealisation of the situation in Figure 9, where the Emission Domain is represented by the emission surface ES, which lies outside the timelike OMOTS surface. Events inside the OMOTS surface are locally trapped. The event horizon lies outside the OMOTS surface; events inside the event horizon are trapped, even if they are not locally trapped; that is, their future necessarily ends up at the future singularity.\\

Denote the event when radiation emission commences as ES(1); it occurs at  time $t_{1}$ in terms of the conformal diagram coordinates (see Figure 9). There will be an event ES(2) at time $t_{2}$ when the ES crosses the event horizon; it lies within the event horizon from then on. \\

The key question is whether ES(1) lies outside the event horizon, or not.
\begin{itemize}
  \item \textbf{Case 1: Bright.} If ES(1) lies outside the event horizon, the situation is as shown in Figure 9. There is an event ES(*) where the ES crosses the event horizon, at time $t_*$. Radiation emitted between times $t_1$ and $t_*$ will escape to infinity. Radiation emitted after time $t_*$ will be trapped, and end up on the singularity. Then define
      \begin{equation}\label{eq:emit}
      \Delta t_{emit} := t_* - t_1 >0.
      \end{equation}
  \item \textbf{Case 2: Dark.} If ES(1) lies inside the event horizon (not shown), almost all the radiation emitted will be trapped, and end up on the singularity. Then define
      \begin{equation}\label{eq:emit1}
      \Delta t_{emit} := 0.
      \end{equation}
\end{itemize}
If $\Delta t_{emit} > 0$, then radiation is emitted to infinity and, as seen from the exterior, this process
never ends:\ collapse is always taking place and radiation is always being emitted by the black hole (albeit its intensity dies away dramatically with time due to redshifting). However the externally perceived mass does not die away to zero. The final mass cannot be affected by the Hawking radiation process at late times ($t> t_*$), when the radiation process  in effect transfers mass from the infalling fluid to the future singularity rather than to infinity. The black hole therefore does not evaporate: it remains in eternal existence (as seen from
the outside) even though the fluid collapsing to a future singularity may get completely radiated away (as seen on the inside). \\

If $\Delta t_{emit} = 0$, then no radiation is emitted to infinity. As seen from the exterior the mass remains constant. The radiation process will indeed be taking place, but it will be an internal process that will not be visible from the exterior. \\

The initial mass $m_{0}$ of the star decreases by the amount of
radiation $\Delta m_{emit}$ emitted in the interval $\Delta
t_{emit}$ if $\Delta t_{emit}\ > 0$ and by zero if $\Delta t_{emit}\
= 0$. The final mass of the black hole is
\begin{equation}  \label{eq:m_final}
m_{final}:=m_{0}-\Delta m_{emit}
\end{equation}%
and the mass measured from the exterior never decreases to less than
this amount, even though the externally received radiation never
comes to an end, as seen from the outside if $\Delta t_{emit}\ > 0$.\\

Of course the real situation is more complex: emission does not take place on a precise surface, but in a local domain about that surface; and in any case the particle concept is suspect. The Emission Domain is centred on the Emission Surface but is larger. Hence in the case labelled Dark, some of the Emission Domain surrounding the Emission Surface will intrude into the domain outside the event horizon, and succeed in emitting radiation that escapes to infinity. But ES(1) would have to be quite close to the event horizon for this to be the case. \\

To find out what the outcome is  will require a full quantum field theory calculation of the radiation stress tensor, taking into account any backreaction effects on the location of the event horizon, that determines if any radiation escapes or not. This will depend on the physics that determines an effective  $\Delta t_{emit}$ and the degree to which the radiation emission is localised around the emission surface and lies beyond the event horizon, and hence determines $\Delta m_{emit}$.

\subsection{Astrophysical Black Holes do not evaporate}
After an initial burst of radiation, mass loss due to the
Hawking radiation process occurs from the collapsing star to the future
singularity FS, not to infinity. The Hawking process thus cannot cause mass
loss from the FS, which rather is partially caused by that radiation.
Most of the mass does not vanish in a pop. It just eternally
sits there in the FS after all the mass of the star has radiated away and
been deposited partly at infinity but mostly in the FS.\\

The final conclusion is that despite a burst of radiation that
may occur, black holes are eternal, when we take Hawking
radiation into account. They may or may not radiate to the outside world; that is a subject for research.
There is in any case a residual mass $m_{final}$ given
by (\ref{eq:m_final}) that never goes away. The mass measured from
the exterior never decreases to less than this amount, even though
any externally received radiation never comes to an end, as seen
from the outside. This conclusion is in agreement with work by Vilkovisky (\cite{Vil06a}-\cite{Vil08}) obtained by very different means. He concludes,
\begin{quote}
\emph{``Black holes create a vacuum matter charge to protect themselves
from the quantum evaporation. A spherically symmetric black hole
having initially no matter charges radiates away about 10\% of the
initial mass and comes to a state in which the vacuum-induced charge
equals the remaining mass.`''
}\end{quote}
This conclusion may well be compatible with what is presented here.\\

The argument in the present paper is based in two key features: firstly the claim that Hawking radiation is emitted locally near timelike MOTS surfaces, as indicated in Figure 5; and secondly the recognition that such surfaces occur at the surface of the collapsing star after the initial closed trapped surface $I_{MOTS}$ forms, as shown in Figure 2. The rest follows from this, and is summarised in Figure 8. Although the context for discussion has been spherically symmetric collapse, the result is likely to be stable for any collapse that leads to the formation of closed trapped surfaces. Thus the argument is based in general relativity considerations, modified by allowing for semiclassical backreaction. It applies to astrophysical black holes, whose horizon will at all times be much greater than the Planck scale, so quantum gravity considerations should not alter the conclusion. \\

Nevertheless there may be counter arguments from the string theory side as follows (I thank Matt Visser for this comment):
\begin{quote}
``\emph{If there are stable remnants, then by standard arguments they must store
a large fraction of the Bekenstein entropy of the black hole they
originated from; but this means they have an enormous number of
internal states;  but then in \emph{any} scattering process virtual
remnants must dominate --- since the propagator is enhanced by a degeneracy factor
$\exp(S_{remnant})$ which will (allegedly) overwhelm any
Planck-scale suppression.}''
\end{quote}
But these states are not accessible to the propagator. They have left spacetime, and so cannot be included in any propagator calculated from within spacetime; that is the essential nature of a singularity, which these arguments do not take seriously. Additionally, this argument requires you to believe standard QFT holds for super-Planck scale black hole remnants. This seems highly questionable. Indeed if it were true, it is not clear why one needs a full quantum gravity theory. At least from a general relativity viewpoint, this counter argument fails to be convincing. The result should hold.

\section{Implications}
\label{sec:option2_Implications}

If this account is correct, it will have implications for various current
debates regarding black holes.

\subsection{Black hole thermodynamics}
\label{sec:option2_Implications_td}

As explained above, the key point is that the ES after a while enters the
event horizon. All later emitted radiation falls into the singularity rather
than going to infinity. Hence there may be a burst of radiation that escapes to
infinity while the ES is outside the event horizon, but the rest does not,
it is trapped. An outside observer may see radiation always being emitted, but
the black hole mass (from the outside) asymptotes to a finite positive
value, as do the temperature and entropy. Inside, the matter can be radiated
away to zero, with no limit on the associated Hawking temperature: but this
radiation does not escape to infinity, it ends up in the singularity The
Hawking radiation is there all the time, and the black hole never goes away.%

\subsubsection{The laws of black hole thermodynamics}
What about the laws of black hole thermodynamics? The area, angular
momentum, and charge of the black hole can be measured at the
spacelike OMOTS surface \cite{AshKri02}. When matter or radiation
falls in, there are changes $\delta A,\, \delta  J,\, \delta Q$
in these quantities. These changes obey the three laws of black hole thermodynamics (\cite{BarCarHaw73}, \cite%
{HawPen96}:23-25):
\begin{itemize}
\item \textit{Second Law}: $\delta A\geq 0$ (\cite{AshKri02}: Section I.)

\item \textit{First law}: $\delta E=\frac{\kappa }{8\pi }\delta A+\Omega
\delta J+\Phi \delta Q$ (\cite{AshKri02}: (20)).

\item \textit{Zeroth Law}: $\kappa $ is the same everywhere on the horizon
of a time independent black hole.
\end{itemize}
The key point here is that one can calculate the mass and angular momentum
of a black hole for the case of an isolated horizon, that is an apparent
horizon: one does not need an event horizon per se. The argument is detailed in \cite{AshKri02} and in \cite{Dreetal03},
where numerical methods for this purpose are given. The OMOTS surface
is accessible to external observers and this is where one can determine the mass
and angular momentum of the black hole \cite{Dreetal03}
(the bad news is that an observer just outside the OMOTS surface will be
inside the event horizon and so won't be able to get that information out; but we won't pursue that issue here).\newline

 These laws will remain unchanged (\cite{AshKri02}, \cite{Nie08}). The Hawking radiation is
there all the time while the black hole settles down to a finite size which
corresponds to a specific final entropy. The new point is that there is a remnant zero point mass
$m_{final}$ when you have got rid of all the disposable energy. One can think of this as an analogue of
all the other zero point energies that occur in quantum theory.\newline

Black holes are paradoxical in that, by (\ref{eq:eos}), as
the entropy goes down, the temperature goes up. If there is a form of the
third law that might hold in the current context of uncharged black holes,
it must hold as regards the entropy rather than the temperature. The issue
in entropy terms is that most thermodynamics is affected by entropy
differences as opposed to absolute entropies \cite{Lem13}; absolute
entropies are relevant only as regards the third law of thermodynamics. On
the standard view, the entropy decays to zero, which is the minimum allowed
value, because the black hole mass and hence surface area decay to zero; but
that is no longer true in the presently considered situation. In the case
envisaged, an uncharged black hole entropy cannot go to zero. This suggests
a new kind of Third Law for this case of uncharged (hence non-extremal)
black holes:

\begin{itemize}
\item \textit{Third law for uncharged black holes}: It is impossible for the
Hawking radiation process to reduce the entropy of a black hole to zero.
\end{itemize}

The value $m_{final}$ that remains determines the
residual entropy of the system, which will be given by (\ref{Eq:S}):
\begin{equation}\label{eq:sfinal}
S_{BH}(final) =4\pi m_{final}^{2}.
\end{equation}
This depends on the value $\Delta m_{emit}$,
which in turn depends on the value $\Delta t_{emit}$
, which follows from the
physics in two ways: firstly, how far outside the MOTS surface is the emission surface ES? Secondly, how
far outside the Initial Null Surface is the event horizon? There may well be
some fundamental formula determining these quantities.

\subsubsection{The quantum case}
An important development in black hole thermodynamics was the
discovery of Hawking radiation just as needed to relate the above
laws to the idea of temperature. It is not clear how the proposal here
relates to that issue, and so to the overall consistency of the
thermodynamic picture when Hawking radiation occurs.\\

Malcolm Perry comments (personal communication) that the problem may be deeper:
\begin{quote}
T\emph{he real Einstein equations are not given by }
\begin{equation*}
\textrm{Einstein tensor = Expectation value of the energy-momentum tensor \,\, (*)}
\end{equation*}
\emph{but rather }
\begin{equation*}
\textrm{Einstein tensor = Energy-Momentum tensor as an operator
expression.
}\end{equation*}
\emph{Thus the gravitational field is the superposition of a
whole collection of things that might well have no relation to
the classical ones. The hydrogen atom being one example. It is a
hope that the semi-classical one (*) is a good and consistent
approximation. It might be, but from what you are saying I suspect
not as I have a hard time ditching the thermodynamic interpretation,
at least in its low temperature limit.}
\end{quote}
I will not pursue that issue here.
\subsection{Astrophysical implications}
\label{sec:option2_Implications_ap}

Permanent astrophysical black holes remain after a finite amount of mass $%
\Delta m_{emit}$ has been radiated away before the star disappears behind
the event horizon (as seen in its comoving frame). Seen from outside, it
will never disappear, and black holes will never dispose of their total mass
into the universe: a mass $m_{final}$ will remain. Thus the amount of
radiation one might expect to detect from black hole evaporation processes
will be less than thought before; indeed it may be very small.\newline

It is the apparent horizon that matters for these purposes, not the event
horizon. In this regard black holes can be treated as local objects (\cite{AshKri02}, \cite%
{Dreetal03}), and you don't have to worry about the far future fate of the
universe in order to know what happens. That is good news. As stated by
Visser \cite{Vis01},
\begin{quote}
``\textit{Remember that to define the event horizon you need to know the
entire history of the spacetime out to the infinite future; you should be a
little alarmed if the question of whether or not a black hole is radiating
now depends on what it is doing in the infinite future.}''
\end{quote}

Luckily this seems not to be the case. In particular, calculations of black hole mergers can be done on a local basis; they do not need to rely on the non-locally determined position of the event horizon. This argument gives substantial support to the main technical point on which this paper is based: that the radiation must be emitted at a dynamical horizon and not the event horizon. \\

Because the evaporation times of astrophysical black holes are so long, the astrophysics of black holes will be unaffected. Their crucial role in many aspects of high energy astrophysics \cite{BegRee95} is unchanged. However Larena and Rothman \cite{LarRot10} point out the remnants of primordial black holes might be back hole candidates. and the detection properties of primordial black holes (\cite{Car05}, \cite{Caretal10}) will be different. The issue to be explored is if the ideas presented here also apply in some form to primordial black holes. That is an issue for future research.
\subsection{Theoretical Implications}
\label{sec:option2_theory}

This viewpoint has obvious implications for the current debates on the
information loss paradox.\newline

Is there information loss? (\cite{Haw76}, \cite{Haw05}, \cite{Mat12}, \cite%
{Gid13}). Yes indeed (\cite{HawPen96}: 59, 63). Gravity is a non-linear
theory with spacetime singularities:\ information falling into them is
inevitably lost. Any matter or information falling in through the event
horizon disappears into the permanent spacelike singularity, which does not
go away and can act as a sink for an arbitrary number of microstates. As no radiation is emitted outwards from the singularity to
infinity (because it is spacelike), no information can be carried out of the
black hole from the singularity by any such radiation. Because there is a
permanent relic, just as in the case Maldacena and Horowitz \cite{HorMal04},
there is no information loss paradox.\newline

Conclusion: Infalling matter and information falls into the
singularity and is destroyed there. It cannot be re-emitted by
Hawking radiation from there as no Hawking radiation from there
reaches infinity. Microstates are swallowed up by the singularity.
From the outside, the black hole acts as an absorbing element, so
scattering off it should not be expected to be unitary (energy will
not be conserved at the event horizon). Additionally, the
astrophysical context indicated here (specifically, the presence of
the CMB radiation) causes rapid decoherence, so entanglement across
the horizon \cite{Mat12} is rapidly lost \cite{Sch07} and associated
problems will therefore dematerialize.\\

The key assumptions leading to this conclusion are,
\begin{itemize}
  \item Hawking radiation is locally rather than globally
  determined; thus radiation emission happen in response to a local
  MOTS rather than the global causal limits associated with an
  event horizon;
  \item The MOTS associated with radiation emission will be timelike, and consequently
  will have a Outward Null Converging (ONC) region to the exterior and an Outward Null Diverging (OND) region to
  the interior. Thus it will be an EMOTS rather than an OMOTS
\end{itemize}
If these assumptions are wrong, the outcome may be different.
\newpage
\section{Conclusion}
\label{sec:conclude}

All this needs to be checked, and further development considered. Specific
issues are,

\begin{itemize}
\item Checking whether the Main Hypothesis (Section \ref{sec:main1}) is indeed correct.

\item Determining if Case 1 (Bright) or  Case 2 (Dark) occurs (Section \ref{sec:mass loss}).

\item Determining what is the value of the remnant mass $m_{final}$, and so final entropy. This
depends on calculating the relation of the Emission Domain to the MOTS surface, and
how much of the Emission Domain remains outside the event horizon, and so determines $\Delta m_{emit}$.

\item Considering other geometries:\ is the result stable to rotation?\ To
perturbed black holes? The nature of the arguments given here
suggest it probably is.

\item If later mass shells fall in, what happens? This should make no
difference: just apply the above argument after the last mass shell has
fallen in.

\item How does this relate to the Hawking evaporation process for primordial black holes? This is unclear, but is obviously worth pursuing. The key question will be how one relates the CMB radiation to primordial black holes.

\item The above analysis assumes the size of the black hole is much greater
than the de Broglie wavelength of the radiation (c.f. \cite{Gid13}). The
situation will clearly be different when that is not true \cite{Mat12}: what
will happen then?

\item The view put here might impact the viewpoint on black holes in a full
quantum gravity theory. That will depend on the theory envisaged.
\end{itemize}
The argument above is based on the broad nature of the processes at work, plus a careful delineation of
the relevant causal domains; detailed calculations of back reaction effects
are necessary in order to confirm this model and determine details of the
outcome. They may lead to a quite different picture; Gary Gibbons comments (private communication)
\begin{quote}
\emph{Much of what you say hinges on the idea
that in QFT  particles exist in a local sense and can be localised
to  spatial or spatio-temporal regions. This certainly seems fishy
in the case of Hawking radiation where, according to the formula $T=
1/8\pi M$  the typical thermal wave length is comparable in size to
the region they originated from. One may at best speak of the expectation value of the stress tensor,
and even for this, dispersion (root means square values) will be
comparable. The particles only emerge when the flux has escaped to
infinity and one has time to localise them. In fact one expects that  even the metric fluctuates, there is no
single metric and  this also makes the idea of localisation suspect.
}
\end{quote}
Readers of this paper should keep that comment in mind in assessing if its results are valid.\\

The bottom line is twofold. Firstly, Figure 7 is much more plausible
the Figure 4; and this is because it takes seriously both the
backreaction effect of the radiation process on the position of the
emission surface and the marginal trapped surfaces (Section
\ref{sec:option2_domains}), and the local nature of that emission
process. The latter is required in order that local astrophysical
calculations of black hole properties make sense (see the Visser
quote in Section \ref{sec:option2_Implications_ap}). \\

Secondly, even if details of what is presented here are unrealistic or the model is dubious, what this discussion shows is that if \emph{any} of the Hawking radiation falls into the singularity, it is unlikely that it can then evaporate away; and then the broad picture presented here will be correct. And even if one changes many details, for instance abandoning a particle model, it is unlikely that \emph{all} the Hawking radiation can avoid a singular fate of this kind. Then the main result will be vindicated.
\newline

\bigskip

\noindent

\textbf{Acknowledgements:}\newline

I gratefully thank Malcolm Perry for enlightening discussions that
initiated this work, and particularly for discussions regarding the
local mechanisms at play in this situation that are embodied in
Figure 5. I thank Reza Tavakol for a very useful reference, Jeff
Murugan, Matt Visser, Paul Davies, Gary Gibbons,and Julien Larena for useful
comments, and Tim Clifton for an important
reference and very helpful discussions that lead to Figure 8.\\

I thank Trinity College, Cambridge, for two terms as a Visiting Fellow
Commoner, and DAMTP, Cambridge, for hospitality. Finally, I acknowledge the
financial support of the University of Cape Town Research Committee and the
National Research Foundation (South Africa).
\newpage

\newpage
\section*{Appendix: The Emission Process}
The emission process can be viewed in particle terms, but it is then necessary to re-express it in terms of the effective energy momentum tensor. The process will be considered here for the case of a timelike EMOTS surface; the null case is degenerate and will not necessarily reflect what happens in a realistic astrophysical context.

\subsection{Particle picture:}
A virtual pair gets converted to a real pair by the intervention of a barrier: in this case, a local trapping horizon (the EMOTS). The left particle tunnels through and becomes real; that causes the other to also become real as the antiparticle of the first, in a version of the EPR paradox (the right hand particle changes when the left hand one tunnels). This second event is outside the horizon.\\

Figure 10a shows this picture. The vacuum is continually emitting virtual particle pairs, but they usually just recombine with no remnant particles left behind. Thus the unconstrained vacuum seethes and bubbles but does not emit particles. This all changes when a separation barrier of some kind is introduced. One particle crosses the barrier and is then unable to return to pair up with its mate; they become real particles. Such emission depends on conditions in a volume $\delta V$ near the trapping surface, so it is not a pointlike process; nevertheless the emission of the pair can be associated with the event E where the virtual pair originated.
\begin{figure}[tbp]
\includegraphics[width=7in]{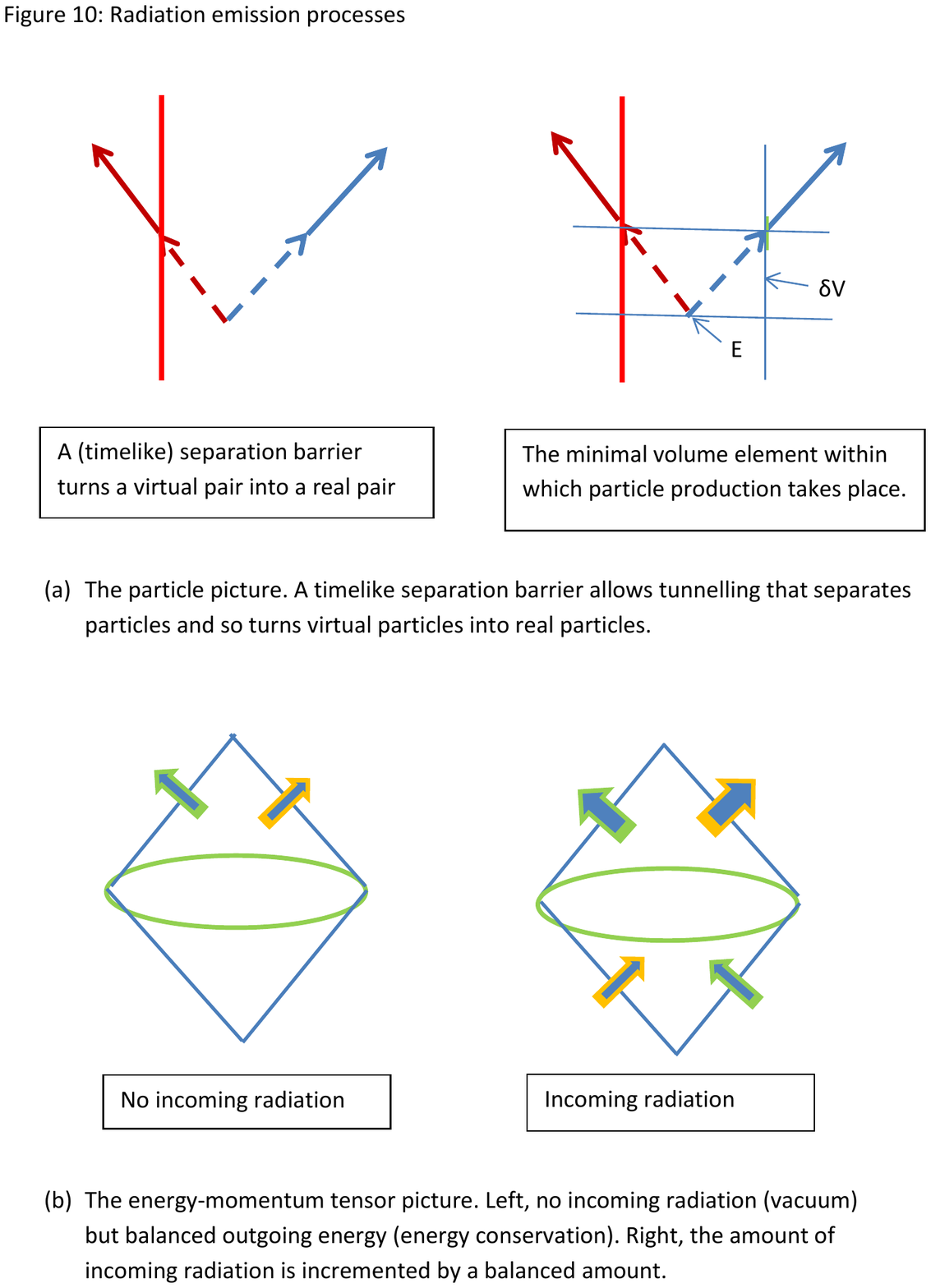}
\caption{Figure 10} \label{Fig10}
\end{figure}
This event is outside the EMOTS, as is the event where the righthand virtual particle pops into existence as a real particle. The volume $\delta V$ is chosen as the smallest volume that encompasses the whole process, as seen form the frame of an observer with 4-velocity chosen to be the timelike eigenvector of the Ricci tensor. This is then the emission domain for that process.

\subsection{Stress tensor version}
This tunneling event is reflected in a bifurcation in the effective stress tensor, which is also a non-local event. \\

 Consider a volume $\Delta V$ locally bounded by future directed and past directed null cones (Figure 10b). This boundary is chosen to be null so that there is no confusion about what is incoming and what is outgoing radiation.  Incoming radiation crosses the lower surface, outgoing radiation crosses the upper surface.\\

In a situation as in Figure 10a where there is initially no incoming radiation (this is the definition of the initial vacuum state), if the process just described takes place within $\Delta V$, this results in opposite outgoing energy flows as judged by the stress tensor. The process is localised within the volume $\Delta V$. \\

In a situation where there is incoming radiation as in Figure 10b, a tunneling process as in Figure 10a will result in enhanced outgoing radiation. The test that an emission event  has taken place is that such a change in the stress tensor occurs. The volume $\Delta V$ should be chosen as the smallest volume in which such a change takes place.\\


\end{document}